\documentclass[11pt]{article}

\usepackage[margin=1in]{geometry}
\usepackage{setspace}
\usepackage{amsmath}
\usepackage{amsfonts}
\usepackage{amssymb}
\usepackage{amsthm}
\usepackage{mathtools}

\usepackage{graphicx}
\usepackage{float}
\usepackage{placeins}

\usepackage{booktabs}
\usepackage{caption}
\usepackage{subcaption}

\usepackage[round]{natbib}

\usepackage{algorithm}
\usepackage{algpseudocode}

\usepackage{xcolor}
\usepackage[colorlinks=true,
            linkcolor=blue,
            citecolor=blue,
            urlcolor=blue]{hyperref}

\theoremstyle{plain}

\theoremstyle{definition}

\theoremstyle{remark}
\newtheorem{remark}{Remark}

\title{Uncertainty-Aware Ideal Point Estimation via Variational EM with P\'{o}lya--Gamma Identity}
\date{} 

\author{
Kwangok Seo\thanks{Department of Statistics, Inha University, Incheon, Korea.}
\and
Youngjo Lee\thanks{Departments of Statistics, Seoul National University, Seoul, Korea.}
\and
Jong Hee Park\thanks{Department of Political Science and International Relations, Seoul National University, Seoul, Korea.}
\and
Xinlei Wang\thanks{Department of Mathematics, University of Texas at Arlington, TX, USA.}
\and
Johan Lim\footnotemark[2]
\thanks{Corresponding author: johanlim@snu.ac.kr}
}
\begin{document}

\maketitle 


\begin{abstract}
    \noindent Roll-call data analysis aims to estimate legislators' ideal points and quantify the associated uncertainty. Existing approaches either rely on Bayesian methods implemented via Markov chain Monte Carlo sampling or focus primarily on point estimation, with uncertainty typically assessed through resampling procedures such as the bootstrap. Consequently, the computational burden of these approaches can become substantial when applied to large roll-call datasets. To address this challenge, we propose a computationally efficient likelihood method for estimating ideal points and their standard errors. Leveraging the P\'{o}lya--Gamma identity, we develop a variational expectation--maximization algorithm for estimating ideal points and introduce a variational Louis' method to approximate the observed Fisher information for standard error estimation. Numerical studies and applications to U.S. congressional roll-call data demonstrate that the proposed method produces accurate ideal point estimates and reliable standard errors while being substantially more computationally efficient than existing approaches.
    
\end{abstract}

\section{Introduction}
Roll-call data record how legislators vote on individual bills, and their quantitative analysis plays a central role in understanding legislative behavior. Two primary objectives are (1) to estimate legislators' ideal points and (2) to quantify the associated uncertainty. 
A variety of approaches have been proposed to address this problem, including \cite{poole1985spatial}, \cite{poole2001geometry}, \cite{clinton2004statistical}, \cite{carroll2013structure}, \cite{imai2016fast}, and \cite{shin2025}. Many of these methods rely on Bayesian estimation via Markov chain Monte Carlo (MCMC) sampling or focus primarily on point estimation, with uncertainty typically assessed through resampling procedures such as the bootstrap \citep{lewis2004measuring, carroll2009measuring, imai2016fast}. Consequently, the computational burden of these approaches can become substantial when applied to large roll-call datasets.

In this paper, we propose a novel likelihood approach for estimating legislators' ideal points together with their associated standard errors.
Our framework begins by modeling the legislative decision-making process within a spatial voting model framework \citep{enelow1984spatial}. Specifically, we consider a quadratic utility function combined with Gumbel-distributed stochastic errors. This formulation gives rise to the two-parameter logistic (2PL) item response theory (IRT) model, in which the parameters correspond to legislators' ideal points and bill-specific parameters.

In our formulation, the ideal points, which constitute the parameters of primary interest, are treated as \emph{fixed} but unknown constants, whereas the bill parameters, regarded as nuisance parameters, are modeled as \emph{random} variables following a Gaussian distribution. 
Under this mixed-effects specification, the objective is to obtain the maximum likelihood estimates (MLEs) of the ideal points along with their corresponding standard errors. 
To achieve this, a plausible approach is to compute the marginal likelihood for the ideal points by integrating out the bill parameters, and then obtain the MLEs and the corresponding observed Fisher information from the resulting marginal likelihood. However, the required integration is analytically intractable and computationally demanding due to the nonlinearity induced by the logistic link. 
As an alternative, one may employ the h-likelihood framework of \cite{lee1996hierarchical}, in which both fixed-effect estimates and their standard errors are obtained via the Laplace approximation \citep{lee2001hierarchical}. However, \cite{jin2024standard} show that the Laplace approximation can yield non-negligible bias in fixed-effect estimation in binary response models and generally fails to provide valid standard errors.

To overcome these difficulties, we employ the P\'{o}lya--Gamma identity \citep{polson2013bayesian}, which provides an augmented likelihood representation that facilitates inference in logistic models. Building on this representation, we develop a variational expectation--maximization (VEM) algorithm \citep{bishop2006pattern} to estimate the ideal points. Using the resulting estimates and variational distributions, we then compute the associated standard errors via the missing information principle \citep{orchard1972missing}, specifically through a variational adaptation of Louis' method \citep{louis1982finding}. The core idea of the missing information principle is that computing the curvature of the marginal likelihood directly is difficult because it requires integrating out the latent variables, but this curvature can instead be recovered from the complete-data log-likelihood.

Extensive numerical experiments demonstrate that the proposed method accurately recovers the true ideal points and yields reliable standard errors. 
In particular, the resulting standard errors closely match those obtained from parametric bootstrap procedures, which are widely regarded as a gold standard, while the computational cost is substantially lower than that of the parametric bootstrap. As an empirical illustration, we apply the proposed method to roll-call data from the 113th and 118th U.S. House of Representatives, demonstrating that it provides a computationally efficient alternative for large-scale roll-call data analysis.

Our proposed method is closely related to that of \cite{cho2021gaussian}. In particular, their approach is similar to ours in that it is based on the 2PL model and treats the parameters of interest as fixed effects while modeling the nuisance parameters as random effects. Consequently, their approach also faces a similar computational challenge in evaluating the marginal likelihood of the parameters of interest. To address this difficulty, they derive a variational lower bound for the extended likelihood \citep{berger1988likelihood} using the Jaakkola--Jordan approximation \citep{jaakkola1997variational}. Treating this lower bound as a complete-data log-likelihood, they develop a VEM algorithm to estimate the parameters of interest (see Section~\ref{sub_sec_2.3} for details).

Although this approach performs well for estimating ideal points, it is not well suited for applying the missing information principle to compute standard errors. The core difficulty is that the curvature of the original extended likelihood can differ substantially from that of the variational lower bound obtained via the Jaakkola--Jordan approximation.

\begin{figure}[!htb]
    \centering
    \includegraphics[width=0.7\linewidth]{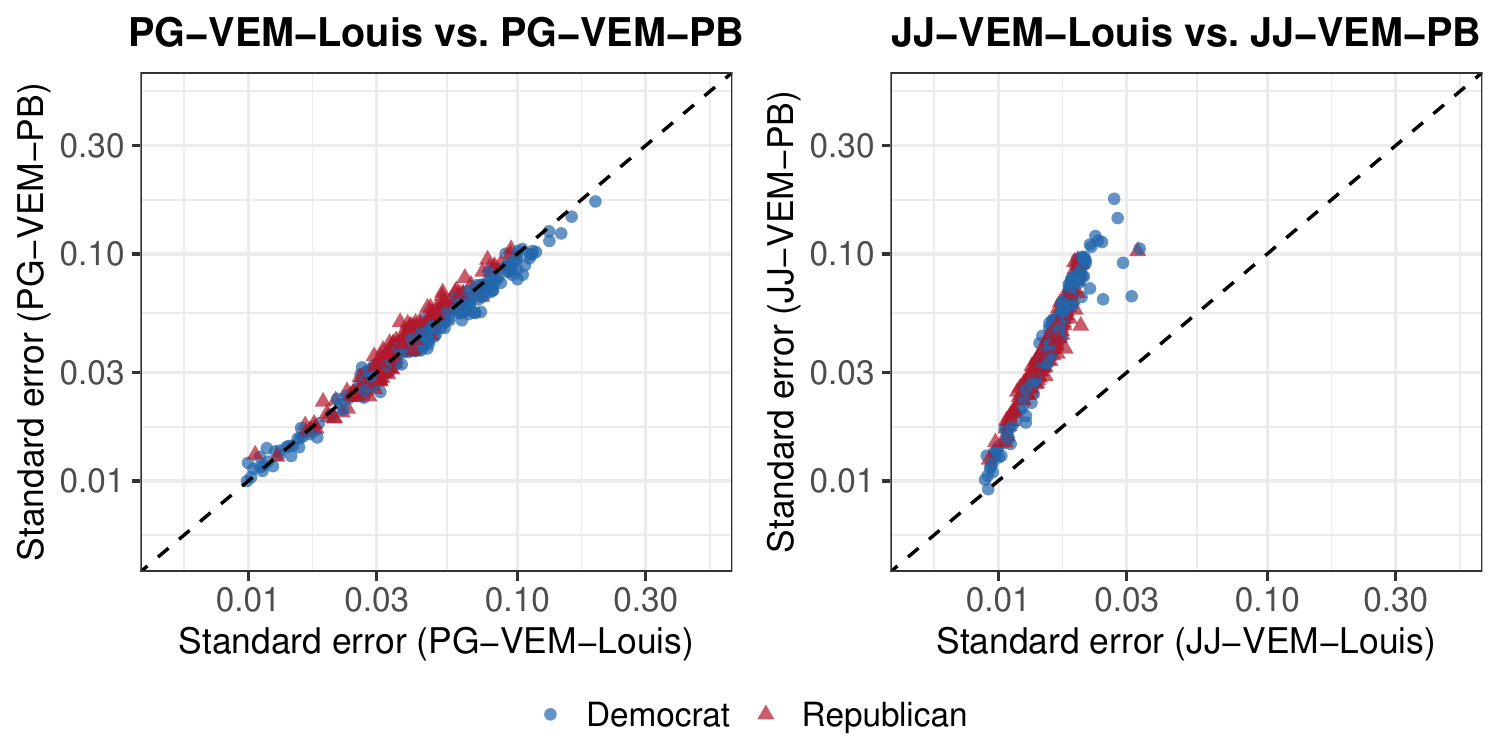}
    \caption{The 112th U.S. Congress: comparison of standard errors for the estimated ideal points. Each panel compares the standard errors from two types of variational Louis' method (\textsc{PG-VEM-Louis} and \textsc{JJ-VEM-Louis}) against their corresponding gold standard parametric bootstrap estimates (\textsc{PG-VEM-PB} and \textsc{JJ-VEM-PB}). Both axes are on a $\log_{10}$ scale, and the black dashed line represents the 45-degree line. Blue circles indicate Democrats, and red triangles indicate Republicans.}
    \label{fig:1}
\end{figure}

Figure~\ref{fig:1} illustrates this limitation using the 112th U.S. Congress dataset. We compare standard error estimates from two different variational Louis' methods---one from the proposed method using the P\'{o}lya--Gamma augmentation (\textsc{PG-VEM-Louis}) and one from the existing method using the Jaakkola--Jordan approximation (\textsc{JJ-VEM-Louis})---against their corresponding parametric bootstrap estimates (\textsc{PG-VEM-PB} and \textsc{JJ-VEM-PB}), which serve as the gold standard. The right panel shows that \textsc{JJ-VEM-Louis} deviates considerably from \textsc{JJ-VEM-PB}, confirming that the Jaakkola--Jordan lower bound is poorly suited for standard error estimation via the variational Louis' method. In contrast, the left panel demonstrates that \textsc{PG-VEM-Louis} closely matches \textsc{PG-VEM-PB}, showing that the P\'{o}lya--Gamma augmentation overcomes this limitation.

\medskip
\noindent\textbf{Outline.} The remainder of this paper is organized as follows.
Section~\ref{sec_2} summarizes the modeling assumptions, the problem setup, and a method closely related to the proposed approach.
Section~\ref{sec_3} presents the proposed method, including a VEM algorithm for estimating ideal points and a variational Louis' method for the associated standard errors. 
Section~\ref{sec_4} presents extensive numerical studies, and Section~\ref{sec_5} provides an empirical application to the 113th U.S. Congress. 
Section~\ref{sec_6} concludes the paper with a brief summary.

\section{Preliminaries}\label{sec_2}

\subsection{Spatial Voting Model}\label{sub_sec_2.1}
Consider a set of $I$ legislators voting on $J$ bills in a legislative body. For each legislator $i$ and bill $j$, we observe a binary vote $Y_{ij}\in\{0,1\}$, where $Y_{ij}=1$ denotes a yea vote and $Y_{ij}=0$ denotes a nay vote. To explain legislators’ voting behavior in an interpretable manner, we adopt a utility-based spatial voting model that links observed votes to latent policy preferences.

Let $\theta_i \in \mathbb{R}$ denote legislator $i$'s ideal point in a one-dimensional policy space. For bill $j$, let $\zeta_j, \psi_j \in \mathbb{R}$ denote the policy positions associated with the yea and nay outcomes, respectively. Given $\theta_i$, $\zeta_j$, and $\psi_j$, we model the utility using a quadratic specification as in \cite{poole2001geometry} and \cite{clinton2004statistical}, with Gumbel-distributed stochastic errors as in \cite{poole1985spatial}. Specifically, the utilities associated with voting yea and nay are defined as
\begin{equation*}
    U_{ij}^{\mathrm{yea}} 
    = 
    -(\theta_i-\zeta_j)^2+\eta_{ij},
    \quad 
    U_{ij}^{\mathrm{nay}} 
    = 
    -(\theta_i-\psi_j)^2+\nu_{ij},
\end{equation*}
where $\eta_{ij}$ and $\nu_{ij}$ are independent stochastic error terms following the Gumbel distribution with common location parameter $0$ and common scale parameter $\sigma_j$. 

Let $\mathrm{Logistic}(\mu,\sigma)$ denote the logistic distribution with location parameter $\mu$ and scale parameter $\sigma$. It can be shown after a simple calculation that $\eta_{ij}-\nu_{ij} \sim \mathrm{Logistic}(0,\sigma_j)$. Combined with the decision rule that legislator $i$ votes yea on bill $j$ whenever $U_{ij}^{\mathrm{yea}} - U_{ij}^{\mathrm{nay}} \ge 0$, it follows that
\begin{equation} \label{eq:2PL}
\begin{split}
    \mathbb{P}\!\left(Y_{ij}=1\right) 
    =
    \mathbb{P}\!\left(U_{ij}^{\mathrm{yea}} - U_{ij}^{\mathrm{nay}}\ge 0 \right)
    =
    \mathbb{P}\!\left( X_{ij} \le \frac{\psi_j^2-\zeta_j^2}{\sigma_j}+\frac{2(\zeta_j-\psi_j)}{\sigma_j}\,\theta_i \right) 
    =
    \sigma(\alpha_j+\beta_j\theta_i), 
\end{split}
\end{equation} 
where $X_{ij}$ is a standard logistic random variable, $\sigma(\cdot)$ denotes its distribution function given by $\sigma(x) = \{1 + \exp(-x)\}^{-1}$, $\alpha_j = (\psi_j^2 - \zeta_j^2)/\sigma_j$, and $\beta_j = 2(\zeta_j - \psi_j)/\sigma_j$. It is worth noting that the spatial voting model in \eqref{eq:2PL} is mathematically equivalent to the 2PL model \citep{birnbaum1968some} in the IRT literature. A more detailed version of the description of the spatial voting model and its related terminology is provided in Appendix~E.

We conclude this section by introducing notation used throughout the remainder of the paper. Let $\tilde{\boldsymbol{\theta}}_i=(1,\theta_i)^\top\in\mathbb{R}^2$ and $\tilde{\boldsymbol{\beta}}_j=(\alpha_j,\beta_j)^\top\in\mathbb{R}^2$, and define $x_{ij}=\alpha_j+\beta_j\theta_i=\tilde{\boldsymbol{\theta}}_i^\top\tilde{\boldsymbol{\beta}}_j$. Let $\boldsymbol{y}=\{y_{11},\ldots,y_{IJ}\}$ denote the observed roll-call outcomes. Denote the collection of ideal points by $\boldsymbol{\Theta}=\{\theta_1,\ldots,\theta_I\}$ and the collection of bill parameters by $\boldsymbol{B}=\{\tilde{\boldsymbol{\beta}}_1,\ldots,\tilde{\boldsymbol{\beta}}_J\}$.

\subsection{Mixed-Effect Model and Problem Setup}\label{sub_sec_2.2}
In the 2PL model in \eqref{eq:2PL}, we treat the ideal points $\theta_i$, the parameters of interest, as fixed effects, whereas the bill parameters $\tilde{\boldsymbol{\beta}}_j$, which are regarded as nuisance parameters, are modeled as random effects satisfying
\begin{equation*}
    \tilde{\boldsymbol{\beta}}_j
    \overset{\mathrm{i.i.d.}}{\sim}
    N_2(\boldsymbol{0},\boldsymbol{\Sigma}_{\tilde{\boldsymbol{\beta}}}),
\end{equation*}
where $N_k(\boldsymbol{\mu},\boldsymbol{\Sigma})$ denotes the $k$-variate normal distribution with mean $\boldsymbol{\mu}$ and covariance matrix $\boldsymbol{\Sigma}$.
Above model differs from commonly used approaches such as Bayesian item response theory (BIRT; \citealp{clinton2004statistical}) and emIRT \citep{imai2016fast} in the following three points: First, BIRT and emIRT are based on a probit link, arising from Gaussian stochastic errors, whereas our spatial voting model leads to a logit link via Gumbel-distributed errors. Second, BIRT and emIRT adopt a fully Bayesian framework in which both the ideal points and bill parameters are modeled as random variables, whereas our approach adopts a mixed-effects formulation in which the ideal points are treated as fixed effects. Third, the point estimators of the ideal points under BIRT and emIRT take the form of posterior summaries, namely the posterior mean or the maximum a posteriori (MAP) estimate, whereas under our approach the ideal points are estimated via maximum likelihood.


Under the standard conditional independence assumption of IRT, the extended likelihood \citep{berger1988likelihood} for the observed voting outcomes in our mixed-effects model is given by
\begin{equation} \label{eq:extended_likelihood}
\begin{split}
    L_e\!\left(
    {\boldsymbol{\Theta}, \boldsymbol{\Sigma}_{\tilde{\boldsymbol{\beta}}}},
    {\boldsymbol{B}};
    \boldsymbol{y}
    \right) 
    &\coloneqq 
    \prod_{i=1}^I \prod_{j=1}^J
    p\!\left(y_{ij}\mid{\theta_i},{\tilde{\boldsymbol{\beta}}_j}\right)
    \prod_{j=1}^J
    p\!\left({\tilde{\boldsymbol{\beta}}_j}\mid{\boldsymbol{\Sigma}_{\tilde{\boldsymbol{\beta}}}}\right) \\
    &= \prod_{i=1}^I \prod_{j=1}^J
    \sigma(x_{ij})^{y_{ij}}
    \big(1-\sigma(x_{ij})\big)^{1-y_{ij}}
    \prod_{j=1}^J
    \phi_2\!\left(
    \tilde{\boldsymbol{\beta}}_j;\boldsymbol{0},\boldsymbol{\Sigma}_{\tilde{\boldsymbol{\beta}}}
    \right).
\end{split}
\end{equation}

\noindent
where $\phi_k(\cdot;\boldsymbol{\mu},\boldsymbol{\Sigma})$ denotes the density of the $k$-variate normal distribution with mean $\boldsymbol{\mu}$ and covariance matrix $\boldsymbol{\Sigma}$. 
Integrating out the random effects $\tilde{\boldsymbol{\beta}}_j$ yields the (marginal) likelihood, which can be expressed as
\begin{equation}\label{eq:marginal_likelihood}
\begin{split}
    L\!\left({\boldsymbol{\Theta}}, {\boldsymbol{\Sigma}_{\tilde{\boldsymbol{\beta}}}};\boldsymbol{y}\right)
    \coloneqq
    \prod_{j=1}^J
    \int
    \left\{
    \prod_{i=1}^I
    \sigma(x_{ij})^{y_{ij}}
    \big(1-\sigma(x_{ij})\big)^{1-y_{ij}}
    \right\}
    \phi_2\!\left(\tilde{\boldsymbol{\beta}}_j;\boldsymbol{0},\boldsymbol{\Sigma}_{\tilde{\boldsymbol{\beta}}}\right)
    \, d\tilde{\boldsymbol{\beta}}_j .
\end{split}
\end{equation}

Our inferential objectives are twofold. First, we seek to obtain estimates of the ideal points $\theta_i$. Second, we quantify the uncertainty associated with the estimated ideal points by providing standard errors. If the marginal likelihood in \eqref{eq:marginal_likelihood} were available in closed form, both objectives could be achieved using classical likelihood theory: the ideal points could be estimated by direct maximization of the marginal likelihood, and standard errors could be obtained from the observed Fisher information. However, the integral in \eqref{eq:marginal_likelihood} does not admit a closed-form expression, which makes achieving our inferential objectives challenging in practice. 

To overcome this challenge, 
\cite{cho2021gaussian} proposed the VEM method in the IRT context, which circumvents the intractable integration in \eqref{eq:marginal_likelihood} via the Jaakkola--Jordan approximation \citep{jaakkola1997variational}. As this method is closely related to ours, we briefly review it in the following section.

\subsection{VEM with the Jaakkola--Jordan Approximation}\label{sub_sec_2.3}
\cite{jaakkola1997variational} showed that the logarithm of the logistic distribution function admits the following lower bound:
\begin{equation}\label{eqn:LB_logit}
\begin{aligned}
    &\log \sigma(x)
    \ge
    \log \sigma(\xi)
    + \frac{x-\xi}{2}
    + \lambda(\xi)(x^2 - \xi^2), 
    \quad
    \lambda(\xi) = \frac{1/2 - \sigma(\xi)}{2\xi},
\end{aligned}
\end{equation}
where $\xi$ is a variational parameter controlling the tightness of the bound. 
By leveraging the lower bound in \eqref{eqn:LB_logit}, we obtain a lower bound for the extended log-likelihood in \eqref{eq:extended_likelihood}:
\begin{align*}
    \ell_e\!\left(
    {\boldsymbol{\Theta}}, {\boldsymbol{\Sigma}_{\tilde{\boldsymbol{\beta}}}}, {\boldsymbol{B}}
    \,;\, \boldsymbol{y}
    \right)
    &\coloneqq
    \log \left\{L_e\!\left(
    {\boldsymbol{\Theta}, \boldsymbol{\Sigma}_{\tilde{\boldsymbol{\beta}}}},
    {\boldsymbol{B}};
    \boldsymbol{y}
    \right)\right\}\\
    &= \sum_{i=1}^I \sum_{j=1}^J
    \Bigg\{
    y_{ij} x_{ij}
    - \log\!\big(1+\exp(x_{ij})\big)
    \Bigg\} 
    + \frac{J}{2} \log \left| \boldsymbol{\Sigma}_{\tilde{\boldsymbol{\beta}}}^{-1} \right|
    - \frac{1}{2} \sum_{j=1}^J
    \tilde{\boldsymbol{\beta}}_j^{\top}
    \boldsymbol{\Sigma}_{\tilde{\boldsymbol{\beta}}}^{-1}
    \tilde{\boldsymbol{\beta}}_j + \mathrm{const.}\\
    &\overset{(\star)}{\ge}
    \sum_{i=1}^I \sum_{j=1}^J
    \Bigg\{
    y_{ij} x_{ij}
    + \log \sigma(\xi_{ij})
    - \frac{x_{ij} + \xi_{ij}}{2}
    + \lambda(\xi_{ij}) \big( x_{ij}^2 - \xi_{ij}^2 \big)
    \Bigg\}\\
    &\qquad
    + \frac{J}{2} \log \left| \boldsymbol{\Sigma}_{\tilde{\boldsymbol{\beta}}}^{-1} \right|
    - \frac{1}{2} \sum_{j=1}^J
    \tilde{\boldsymbol{\beta}}_j^{\top}
    \boldsymbol{\Sigma}_{\tilde{\boldsymbol{\beta}}}^{-1}
    \tilde{\boldsymbol{\beta}}_j + \mathrm{const.} \\
    &\eqqcolon
    \ell_e^{\mathrm{JJ}}\!\left(
    {\boldsymbol{\Theta}}, {\boldsymbol{\Sigma}_{\tilde{\boldsymbol{\beta}}}}, {\boldsymbol{\xi}}, {\boldsymbol{B}}
    \,;\, \boldsymbol{y}
    \right).
\end{align*}
where $\boldsymbol{\xi} = \{\xi_{11}, \ldots, \xi_{IJ}\}$. The variational parameters $\xi_{ij}$ control the tightness of the inequality $(\star)$ and are treated as fixed but unknown constants.

To obtain estimates of the ideal points, one can treat the lower bound $\ell_e^{\mathrm{JJ}}$ as a complete-data log-likelihood and then apply the EM algorithm. We refer to this method as \textsc{JJ-VEM}. Relative to the original extended log-likelihood $\ell_e$, the lower bound $\ell_e^{\mathrm{JJ}}$ leads to a more tractable E-step, as the first term of $\ell_e^{\mathrm{JJ}}$ is quadratic in $\tilde{\boldsymbol{\beta}}_j$. \textsc{JJ-VEM} may be viewed as a modified version of \cite{cho2021gaussian} in which the roles of the fixed and random effects are interchanged to better align with the context of roll-call data analysis. Details of the corresponding VEM algorithm are given in Appendix~\ref{appendix_A}.

\textsc{JJ-VEM} performs well in estimating the ideal points; however, it does not provide standard errors for the resulting estimates. To obtain standard errors, one may employ a parametric bootstrap procedure, as adopted in \cite{lewis2004measuring}, \cite{carroll2009measuring}, and \cite{imai2016fast}. Such bootstrap-based inference can be computationally demanding, particularly as the size of the roll-call data increases. More recently, \cite{xiao2024note} proposed applying the missing information principle \citep{orchard1972missing} to compute standard errors under the \textsc{JJ-VEM} framework. However, the missing information principle fundamentally relies on the curvature of the true complete-data log-likelihood $\ell_e$ in order to recover the observed Fisher information. Within the \textsc{JJ-VEM} framework, the true extended log-likelihood $\ell_e$ is replaced by its variational lower bound $\ell_e^{\mathrm{JJ}}$, which can have substantially different curvature. Consequently, applying missing information principle to the surrogate $\ell_e^{\mathrm{JJ}}$ rather than to $\ell_e$ can yield a biased approximation of the observed Fisher information, resulting in unreliable standard errors. Indeed, the simulation results of \cite{xiao2024note} confirm that the standard errors obtained in this manner systematically deviate from those produced by the parametric bootstrap.

A natural research question is whether it is possible to obtain ideal point estimates and their standard errors in a computationally efficient manner without compromising estimation accuracy. Section~\ref{sec_3} provides an affirmative answer to this question.

\section{Proposed Method}\label{sec_3}
In Section~\ref{sub_sec_3.1}, we describe a VEM method based on the P\'{o}lya--Gamma identity for estimating the ideal points. 
In Section~\ref{sub_sec_3.2}, we propose a variational Louis' method that approximates the observed Fisher information using the ideal point estimates together with the variational distributions obtained from the VEM procedure.

\subsection{VEM with the P\'{o}lya--Gamma Identity}\label{sub_sec_3.1}

We begin by briefly introducing the P\'olya--Gamma distribution. A random variable $W$ follows a P\'olya--Gamma distribution with parameters $b>0$ and $c \in \mathbb{R}$,
denoted $W \sim \mathrm{PG}(b,c)$, if
\[
    W \stackrel{d}{=} \frac{1}{2\pi^2} \sum_{k=1}^{\infty} \frac{g_k}{(k-1/2)^2 +
    c^2/(4\pi^2)},
\]
where each $g_k \sim \mathrm{Ga}(b,1)$ is an independent gamma random variable. 
There are two key properties of this distribution that are used later. First, the first moment of $W$ is available in closed form:
\[
    \mathbb{E}[W] = \frac{b}{2c}\tanh(c/2), \qquad W \sim \mathrm{PG}(b,c).
\]
Second, the distribution satisfies an exponential tilting property: if $W \sim \mathrm{PG}(b,0)$, then tilting its density by $\exp(-\tfrac{1}{2}c^2 w)$ and renormalizing yields the density of $\mathrm{PG}(b,c)$, that is,
\[
    \mathrm{PG}(w; b,c) = \frac{\exp(-\tfrac{1}{2}c^2 w)\, \mathrm{PG}(w; b,0)}
    {\mathbb{E}\!\left[\exp(-\tfrac{1}{2}c^2 W)\right]},
\]
where $\mathrm{PG}(\cdot\,;\,b,c)$ denotes the density of a
P\'{o}lya--Gamma random variable with parameters $(b,c)$, and the expectation in the denominator is taken with respect to $W \sim \mathrm{PG}(b,0)$.

We now introduce the P\'{o}lya--Gamma identity described in Theorem~1 of \cite{polson2013bayesian}, which states that for $b>0$ and all $a \in \mathbb{R}$, the following identity holds:
\begin{equation}\label{eq:PG_identity}
\begin{split}
    \frac{(e^{x})^{a}}{(1+e^{x})^{b}}
    =
    2^{-b} \exp\!\left(\left(a-\frac{b}{2}\right)x\right)
    \int_{0}^{\infty}
    \exp\!\left(-\frac{w x^{2}}{2}\right)
    \,\mathrm{PG}(w; b,0)\,dw.
\end{split}
\end{equation}
Applying the identity in \eqref{eq:PG_identity} with $a=y_{ij}$ and $b=1$, the extended likelihood in \eqref{eq:extended_likelihood} admits the following integral representation:
\begin{align*}
    L_e\!\left(
    {\boldsymbol{\Theta}}, {\boldsymbol{\Sigma}_{\tilde{\boldsymbol{\beta}}}},
    {\boldsymbol{B}}; \boldsymbol{y}
    \right)
    &=
    \prod_{i=1}^I \prod_{j=1}^J
    \frac{\exp(x_{ij}y_{ij})}{1+\exp(x_{ij})}
    \prod_{j=1}^J
    \phi_2\!\left(
    \tilde{\boldsymbol{\beta}}_j;\boldsymbol{0},\boldsymbol{\Sigma}_{\tilde{\boldsymbol{\beta}}}
    \right)\\
    &=
    \prod_{i=1}^I \prod_{j=1}^J
    \int_{0}^{\infty}
    \frac{1}{2}
    \exp\!\Bigg(
    \left(y_{ij}-\frac{1}{2}\right)x_{ij}
    -\frac{1}{2}w_{ij}x_{ij}^2
    \Bigg)\mathrm{PG}(w_{ij};1,0)\,dw_{ij}
    \prod_{j=1}^J
    \phi_2\!\left(
    \tilde{\boldsymbol{\beta}}_j;\boldsymbol{0},\boldsymbol{\Sigma}_{\tilde{\boldsymbol{\beta}}}
    \right)\\
    &=
    \int_{0}^{\infty}\cdots\int_{0}^{\infty}
    \prod_{i=1}^I \prod_{j=1}^J
    \frac{1}{2}
    \exp\!\Bigg(
    \left(y_{ij}-\frac{1}{2}\right)x_{ij}
    -\frac{1}{2}w_{ij}x_{ij}^2
    \Bigg)
    \mathrm{PG}(w_{ij};1,0)\\
    &\qquad \times
    \prod_{j=1}^J
    \phi_2\!\left(
    \tilde{\boldsymbol{\beta}}_j;\boldsymbol{0},\boldsymbol{\Sigma}_{\tilde{\boldsymbol{\beta}}}
    \right)
    \,dw_{11}\cdots dw_{IJ}.
\end{align*}
Denote the integrand of the final expression by
\begin{align*}
    L_e^{\mathrm{PG}}\!\left({\boldsymbol{\Theta}}, {\boldsymbol{\Sigma}_{\tilde{\boldsymbol{\beta}}}}, {\boldsymbol{w}}, {\boldsymbol{B}}; \boldsymbol{y}\right)
    \coloneqq
    \prod_{i=1}^I \prod_{j=1}^J
    \frac{1}{2}
    \exp\!\Bigg(
    \left(y_{ij}-\frac{1}{2}\right)x_{ij}
    -\frac{1}{2}w_{ij}x_{ij}^2
    \Bigg)
    \mathrm{PG}(w_{ij};1,0)
    \prod_{j=1}^J
    \phi_2\!\left(
    \tilde{\boldsymbol{\beta}}_j;\boldsymbol{0},\boldsymbol{\Sigma}_{\tilde{\boldsymbol{\beta}}}
    \right),
\end{align*}
where $\boldsymbol{w}=\{w_{11},\ldots,w_{IJ}\}$, which are treated as latent random effects.
By taking the logarithm of $L_e^{\mathrm{PG}}$, we obtain
\begin{align*}
    \ell_e^{\mathrm{PG}}\!\left({\boldsymbol{\Theta}}, {\boldsymbol{\Sigma}_{\tilde{\boldsymbol{\beta}}}}, {\boldsymbol{w}}, {\boldsymbol{B}}; \boldsymbol{y}\right)
    &\coloneqq
    \log\!\Bigg(
    L_e^{\mathrm{PG}}\!\left({\boldsymbol{\Theta}}, {\boldsymbol{\Sigma}_{\tilde{\boldsymbol{\beta}}}}, {\boldsymbol{w}}, {\boldsymbol{B}}; \boldsymbol{y}\right)
    \Bigg)\\
    &=
    \sum_{i=1}^I \sum_{j=1}^J
    \Bigg\{
    \log(1/2)
    +\left(y_{ij}-\frac{1}{2}\right)x_{ij}
    -\frac{1}{2}w_{ij}x_{ij}^2
    +\log\bigg(\mathrm{PG}(w_{ij};1,0)\bigg)
    \Bigg\}\\
    &\qquad
    +\frac{J}{2}\log\left|\boldsymbol{\Sigma}_{\tilde{\boldsymbol{\beta}}}^{-1}\right|
    -\frac{1}{2}\sum_{j=1}^J
    \tilde{\boldsymbol{\beta}}_j^{\top}
    \boldsymbol{\Sigma}_{\tilde{\boldsymbol{\beta}}}^{-1}
    \tilde{\boldsymbol{\beta}}_j + \mathrm{const.}
\end{align*}
It is important to note that $L_e^{\mathrm{PG}}$ specifies a joint density over the observed votes $\boldsymbol{y}$ and the latent variables $\boldsymbol{w}$ and $\boldsymbol{B}$: integrating $L_e^{\mathrm{PG}}$ over $\boldsymbol{w}$ recovers the extended likelihood $L_e$ in \eqref{eq:extended_likelihood}, and further integrating over $\boldsymbol{B}$ recovers the marginal likelihood $L$ in \eqref{eq:marginal_likelihood}. Motivated by this fact, to estimate the ideal points, we treat $\ell_e^{\mathrm{PG}}$ as a complete-data log-likelihood and develop a VEM algorithm.
The proposed VEM algorithm consists of two steps: a variational E-step and an M-step. The $(t+1)$th iteration of the VEM algorithm proceeds as follows:

\medskip
\noindent\textbf{Variational E-step.}\\
Let $\boldsymbol{z} = \{\boldsymbol{w}, \boldsymbol{B}\}$ denote the collection of latent variables, and let $\boldsymbol{\vartheta} = \{\boldsymbol{\Theta}, \boldsymbol{\Sigma}_{\tilde{\boldsymbol{\beta}}}\}$ denote the model parameters. The complete-data likelihood is given by $L_e^{\mathrm{PG}}(\boldsymbol{\vartheta}, \boldsymbol{z}; \boldsymbol{y}) = p(\boldsymbol{y}, \boldsymbol{z}; \boldsymbol{\vartheta}).$ 
Since the conditional distribution $p(\boldsymbol{z} \mid \boldsymbol{y}; \boldsymbol{\vartheta})$ is intractable, we employ variational inference (VI) \citep{jordan1999introduction, blei2017variational} to approximate it.
Specifically, given the current parameter estimates $\boldsymbol{\vartheta}^{(t)}$, we seek a variational distribution $q^{*(t+1)}(\boldsymbol{z})$ that minimizes the Kullback--Leibler divergence to the conditional distribution $p(\boldsymbol{z}\mid \boldsymbol{y}; \boldsymbol{\vartheta}^{(t)})$ over the prespecified variational family $\mathcal{Q}$:
\begin{equation}\label{eq:KL-div}
    q^{*(t+1)}(\boldsymbol{z})
    =
    \underset{q(\boldsymbol{z}) \in \mathcal{Q}}{\arg\min}
    \;
    \mathrm{KL}\!\left(
    q(\boldsymbol{z})
    \,\middle\|\,
    p(\boldsymbol{z} \mid \boldsymbol{y}; \boldsymbol{\vartheta}^{(t)})
    \right).
\end{equation}
In this paper, we consider a mean-field variational family $\mathcal{Q}$ of the form
\begin{equation*}
    \mathcal{Q}
    \coloneqq
    \left\{
    q(\boldsymbol{z})
    =
    \prod_{i=1}^I \prod_{j=1}^J q_{ij}(w_{ij})
    \prod_{j=1}^J q_j(\tilde{\boldsymbol{\beta}}_j)
    \right\},
\end{equation*}
which imposes independence among the P\'{o}lya--Gamma latent variables $w_{ij}$ and the bill parameters $\tilde{\boldsymbol{\beta}}_j$.

The optimization problem in \eqref{eq:KL-div} can be equivalently reformulated in terms of the evidence lower bound (ELBO). Specifically, minimizing the Kullback--Leibler divergence is equivalent to maximizing the ELBO, defined as
\begin{equation*}
\begin{split}
    \mathrm{ELBO}(q(\boldsymbol{z}); \boldsymbol{\vartheta}^{(t)})\coloneqq
    \mathbb{E}_{q(\boldsymbol{z})}[\log p(\boldsymbol{y},\boldsymbol{z};\boldsymbol{\vartheta}^{(t)})]
    -
    \mathbb{E}_{q(\boldsymbol{z})}[\log q(\boldsymbol{z})].
\end{split}
\end{equation*}
This equivalence follows from the decomposition
\begin{equation*}
\begin{split}
    \mathrm{KL}\!\left(
    q(\boldsymbol{z})
    \,\middle\|\,
    p(\boldsymbol{z}\mid \boldsymbol{y};\boldsymbol{\vartheta}^{(t)})
    \right)
    =
    -\mathrm{ELBO}(q(\boldsymbol{z}); \boldsymbol{\vartheta}^{(t)})
    +
    \log p(\boldsymbol{y};\boldsymbol{\vartheta}^{(t)}),
\end{split}
\end{equation*}
where $\log p(\boldsymbol{y};\boldsymbol{\vartheta}^{(t)})$ does not depend on $q(\boldsymbol{z})$. Consequently, the optimization problem in \eqref{eq:KL-div} can be equivalently written as
\begin{equation}\label{eq:ELBO}
    q^{*(t+1)}(\boldsymbol{z})
    =
    \underset{q(\boldsymbol{z})\in\mathcal{Q}}{\arg\max}
    \;\mathrm{ELBO}(q(\boldsymbol{z}); \boldsymbol{\vartheta}^{(t)}).
\end{equation}

The optimization problem in \eqref{eq:ELBO} can be solved via coordinate ascent variational inference (CAVI, \citealp{bishop2006pattern, blei2017variational}), which iteratively optimizes each factor of the mean-field variational distribution while holding the others fixed. Without loss of generality, we first update the factors corresponding to the P\'{o}lya--Gamma variables $w_{ij}$ and then those of the bill parameters $\tilde{\boldsymbol{\beta}}_j$.

Note that the proposed VEM algorithm consists of two nested iterative procedures: an outer EM algorithm indexed by $t$ and an inner CAVI algorithm indexed by $s$. At the $(t+1)$th EM iteration, the $(s+1)$th iteration of CAVI proceeds as follows:

\medskip
\noindent\emph{Update for $q_{ij}(w_{ij})$. }\\
Holding $\{q_{11}^{(t+1,s+1)}(w_{11})$, $\ldots$, $q_{ij-1}^{(t+1,s+1)}(w_{ij-1})$, $q_{ij+1}^{(t+1,s)}(w_{ij+1})$, $\ldots$, $q_{IJ}^{(t+1,s)}(w_{IJ})\}$ and $\{q_1^{(t+1,s)}(\tilde{\boldsymbol{\beta}}_1)$, $\ldots$, $q_J^{(t+1,s)}(\tilde{\boldsymbol{\beta}}_j)\}$ fixed, the CAVI update for $q_{ij}(w_{ij})$ is obtained by maximizing $\mathrm{ELBO}(q(\boldsymbol{z});\boldsymbol{\vartheta}^{(t)})$ with respect to $q_{ij}(w_{ij})$. Under the mean-field approximation and by the law of iterated expectation, the ELBO can be written, up to an additive constant independent of $q_{ij}(w_{ij})$, as
\begin{equation*}
\begin{split}
    \mathrm{ELBO}(q_{ij}(w_{ij}); \boldsymbol{\vartheta}^{(t)})
    &=
    \mathbb{E}_{q_{ij}(w_{ij})}
    \Bigg[
    \mathbb{E}_{q_j^{(t+1,s)}(\tilde{\boldsymbol{\beta}}_j)}
    \Bigg[
    -\frac{1}{2}w_{ij}\left(x_{ij}^{(t)}\right)^2
    +\log\bigg(\mathrm{PG}(w_{ij};1,0)\bigg)
    \Bigg]
    \Bigg]\\
    &\qquad
    -\mathbb{E}_{q_{ij}(w_{ij})}[\log q_{ij}(w_{ij})]
    +\mathrm{const.}
\end{split}
\end{equation*}
Maximizing this expression is equivalent to minimizing the KL divergence between $q_{ij}(w_{ij})$ and a distribution proportional to
\begin{equation*}
\begin{split}
    \exp\!\Bigg(
    \mathbb{E}_{q_j^{(t+1,s)}(\tilde{\boldsymbol{\beta}}_j)}
    \Bigg[
    -\frac{1}{2}w_{ij}\left(x_{ij}^{(t)}\right)^2
    +\log\bigg(\mathrm{PG}(w_{ij};1,0)\bigg)
    \Bigg]
    \Bigg),
\end{split}
\end{equation*}
which yields
\begin{equation*}
\begin{split}
    q_{ij}^{(t+1,s+1)}(w_{ij})
    &\propto
    \exp\!\Bigg(
    \mathbb{E}_{q_j^{(t+1,s)}(\tilde{\boldsymbol{\beta}}_j)}
    \Bigg[
    -\frac{1}{2}w_{ij}\left(x_{ij}^{(t)}\right)^2
    +\log\bigg(\mathrm{PG}(w_{ij};1,0)\bigg)
    \Bigg]
    \Bigg)\\
    &=
    \exp\!\left(
    -\frac{1}{2}w_{ij}\,
    \mathbb{E}_{q_j^{(t+1,s)}(\tilde{\boldsymbol{\beta}}_j)}\left[\left(x_{ij}^{(t)}\right)^2\right]
    \right)
    \mathrm{PG}(w_{ij};1,0).
\end{split}
\end{equation*}
By the exponential tilting property of the P\'{o}lya--Gamma distribution, this leads to the closed-form update

\begin{equation}\label{eq:wij_CAVI}
\begin{split}
    q_{ij}^{(t+1,s+1)}(w_{ij})
    =
    \mathrm{PG}\!\left(1,\gamma_{ij}^{(t+1,s+1)}\right),
    \quad \bigl(\gamma_{ij}^{(t+1,s+1)}\bigr)^2
    \coloneqq
    \mathbb{E}_{q_j^{(t+1,s)}(\tilde{\boldsymbol{\beta}}_j)}\!\left[\left(x_{ij}^{(t)}\right)^2\right].
\end{split}
\end{equation}

\medskip
\noindent\emph{Update for $q_j(\tilde{\boldsymbol{\beta}}_j)$. }\\
Holding $\{q_1^{(t+1,s)}(\tilde{\boldsymbol{\beta}}_1)$, $\ldots$, $q_{j-1}^{(t+1,s+1)}(\tilde{\boldsymbol{\beta}}_{j-1})$, $q_{j+1}^{(t+1,s)}(\tilde{\boldsymbol{\beta}}_{j+1})$, $\ldots$, $q_J^{(t+1,s)}(\tilde{\boldsymbol{\beta}}_J)\}$ and $\{q_{11}^{(t+1,s+1)}(w_{11})$, $\ldots$, $q_{IJ}^{(t+1,s+1)}(w_{IJ})\}$ fixed, the CAVI update for $q_j(\tilde{\boldsymbol{\beta}}_j)$ is obtained by maximizing $\mathrm{ELBO}(q(\boldsymbol{z});\boldsymbol{\vartheta}^{(t)})$ with respect to $q_j(\tilde{\boldsymbol{\beta}}_j)$. By analogous calculations to those used for the update of $q_{ij}(w_{ij})$, we obtain
\begin{equation}\label{eq:beta_tilde_j_CAVI}
    q_j^{(t+1, s+1)}(\tilde{\boldsymbol{\beta}}_j)
    =
    N\!\left(
    \tilde{\boldsymbol{\beta}}_j \mid
    \tilde{\boldsymbol{\mu}}_j^{(t+1,s+1)},\, \tilde{\boldsymbol{\Sigma}}_j^{(t+1,s+1)}
    \right),
\end{equation}
where
\begin{align*}
    \tilde{\boldsymbol{\Sigma}}_j^{(t+1,s+1)}
    &=
    \left\{
    \left(\boldsymbol{\Sigma}_{\tilde{\boldsymbol{\beta}}}^{(t)}\right)^{-1}
    +
    \sum_{i=1}^I \bar{w}_{ij}^{(t+1,s+1)}
    \tilde{\boldsymbol{\theta}}_i^{(t)} \tilde{\boldsymbol{\theta}}_i^{(t)\top}
    \right\}^{-1},\\
    \tilde{\boldsymbol{\mu}}_j^{(t+1,s+1)}
    &=
    \tilde{\boldsymbol{\Sigma}}_j^{(t+1,s+1)}
    \left\{
    \sum_{i=1}^I \left(y_{ij}-\frac{1}{2}\right)\tilde{\boldsymbol{\theta}}_i^{(t)}
    \right\},
\end{align*}
and $\bar{w}_{ij}^{(t+1,s+1)} \coloneqq \mathbb{E}_{q_{ij}^{(t+1,s+1)}(w_{ij})}[w_{ij}] = \tanh(\gamma_{ij}^{(t+1,s+1)}/2)/(2\,\gamma_{ij}^{(t+1,s+1)})$.

We iterate the above CAVI updates until convergence under a prespecified stopping criterion, and denote the resulting variational distributions by $q^{(t+1)}(\boldsymbol{z}) = \prod_{i = 1}^I \prod_{j = 1}^J q_{ij}^{(t+1)}(w_{ij}) \prod_{j = 1}^J q_j^{(t+1)}(\tilde{\boldsymbol{\beta}}_j)$.

\bigskip
\noindent\textbf{M-step.}\\
The goal is to maximize the $Q$-function defined as
\begin{align*}
    Q(\boldsymbol{\vartheta}\mid \boldsymbol{\vartheta}^{(t)})
    &\coloneqq
    \mathbb{E}_{q^{(t+1)}(\boldsymbol{z})}
    \!\left[
    \ell_e^{\mathrm{PG}}\!\left(\boldsymbol{\vartheta}, \boldsymbol{z}; \boldsymbol{y}\right)
    \right]\\
    &= \sum_{i=1}^I \sum_{j=1}^J
    \Bigg\{
    \log(1/2)
    +\left(y_{ij}-\frac{1}{2}\right)\tilde{\boldsymbol{\theta}}_i^\top \tilde{\boldsymbol{\mu}}_j^{(t+1)}
    -\frac{1}{2}\bar{w}_{ij}^{(t+1)}\tilde{\boldsymbol{\theta}}_i^\top
    \left(\tilde{\boldsymbol{\Sigma}}^{(t+1)} + \tilde{\boldsymbol{\mu}}_j^{(t+1)}\tilde{\boldsymbol{\mu}}_j^{(t+1) \top} \right)
    \tilde{\boldsymbol{\theta}}_i
    \Bigg\}\\
    &\qquad
    + \sum_{i=1}^I \sum_{j=1}^J \mathbb{E}_{q_{ij}^{(t+1)}(w_{ij})}\left[\log\bigg(\mathrm{PG}(w_{ij};1,0)\bigg)\right]\\
    &\qquad
    +\frac{J}{2}\log\left|\boldsymbol{\Sigma}_{\tilde{\boldsymbol{\beta}}}^{-1}\right|
    -\frac{1}{2}\sum_{j=1}^J
    \mathrm{tr}\left(\boldsymbol{\Sigma}_{\tilde{\boldsymbol{\beta}}}^{-1}\left(\tilde{\boldsymbol{\Sigma}}_j^{(t+1)} + \tilde{\boldsymbol{\mu}}_j^{(t+1)}\tilde{\boldsymbol{\mu}}_j^{(t+1)\top}\right)\right) + \mathrm{const.},
\end{align*}
with respect to $\boldsymbol{\vartheta}$. Here, $\mathrm{tr}(A)$ denotes the trace of the matrix $A$. By setting the gradient of the $Q$-function with respect to $\theta_i$ and $\Sigma_{\tilde{\boldsymbol{\beta}}}$ to zero, we obtain the closed-form update rules
\begin{equation}\label{eq:theta_i_EM}
\begin{split}
    \theta_i^{(t+1)}
    = 
    \frac{\sum_{j=1}^J \Bigg\{\left(y_{ij}-\frac{1}{2}\right)\,(\tilde{\boldsymbol{\mu}}_j^{(t+1)})_{(2)}
    -\left(\tilde{\boldsymbol{\Sigma}}_j^{(t+1)}+
    \tilde{\boldsymbol{\mu}}_j^{(t+1)}\tilde{\boldsymbol{\mu}}_j^{(t+1)\top}\right)_{(1,2)}\Bigg\}}{\sum_{j=1}^J \bar{w}_{ij}^{(t+1)}
    \left(\tilde{\boldsymbol{\Sigma}}_j^{(t+1)}+
    \tilde{\boldsymbol{\mu}}_j^{(t+1)}\tilde{\boldsymbol{\mu}}_j^{(t+1)\top}\right)_{(2,2)}},
\end{split}
\end{equation}
\begin{equation}\label{eq:Sigma_EM}
    \boldsymbol{\Sigma}_{\tilde{\boldsymbol{\beta}}}^{(t+1)}
    =
    \frac{1}{J}\sum_{j=1}^J
    \left(\tilde{\boldsymbol{\Sigma}}_j^{(t+1)}+
    \tilde{\boldsymbol{\mu}}_j^{(t+1)}\tilde{\boldsymbol{\mu}}_j^{(t+1)\top}\right),
\end{equation}   
where $(\cdot)_{(a)}$ denotes the $a$th component of a vector and $(\cdot)_{(a,b)}$ denotes the $(a,b)$th entry of a matrix. The proposed VEM procedure is summarized in Algorithm~\ref{alg:PG_MFVEM}. We refer to this method as \textsc{PG-VEM}.

\begin{algorithm}[!hbt]
\caption{Variational EM via the P\'{o}lya--Gamma Identity}
\label{alg:PG_MFVEM}
\begin{algorithmic}[1]

\Require Observed roll-call data $\boldsymbol{y}$

\Ensure Estimates $\boldsymbol{\Theta}$ and $\boldsymbol{\Sigma}_{\tilde{\boldsymbol{\beta}}}$

\State Initialize $\boldsymbol{\vartheta}^{(0)}=\{\boldsymbol{\Theta}^{(0)},\boldsymbol{\Sigma}_{\tilde{\boldsymbol{\beta}}}^{(0)}\}$ and $\{q_j^{(0)}(\tilde{\boldsymbol{\beta}}_j)\}$
\State Set $t \gets 0$

\Repeat \Comment{Outer EM iteration}

\Statex \textbf{Variational E-step (given $\boldsymbol{\vartheta}^{(t)}$)}

\State Warm-start: set $q_j^{(t+1,0)}(\tilde{\boldsymbol{\beta}}_j) \gets q_j^{(t)}(\tilde{\boldsymbol{\beta}}_j)$ for $j=1,\dots,J$
\State Set $s \gets 0$

\Repeat \Comment{Inner CAVI iteration}

\State Update $\{q_{ij}^{(t+1,s+1)}(w_{ij})\}_{i,j}$ using Eq.~\eqref{eq:wij_CAVI}

\State Update $\{q_j^{(t+1,s+1)}(\tilde{\boldsymbol{\beta}}_j)\}_{j}$ using Eq.~\eqref{eq:beta_tilde_j_CAVI}

\State $s \gets s+1$
\Until{CAVI stopping criterion is satisfied}

\State Set $q_{ij}^{(t+1)}(w_{ij}) \gets q_{ij}^{(t+1,s)}(w_{ij})$ and $q_j^{(t+1)}(\tilde{\boldsymbol{\beta}}_j) \gets q_j^{(t+1,s)}(\tilde{\boldsymbol{\beta}}_j)$

\Statex \textbf{M-step}

\State Update $\{\theta_i^{(t+1)}\}$ using Eq.~\eqref{eq:theta_i_EM}

\State Update $\boldsymbol{\Sigma}_{\tilde{\boldsymbol{\beta}}}^{(t+1)}$ using Eq.~\eqref{eq:Sigma_EM}

\State $t \gets t+1$
\Until{EM stopping criterion is satisfied}

\end{algorithmic}
\end{algorithm}

\begin{remark}
    In principle, the inner CAVI loop can be terminated once the change in the variational parameters falls below a prespecified tolerance, or once a prespecified maximum number of iterations is reached. In practice, however, we find that performing a single CAVI update per EM iteration before proceeding to the M-step already yields accurate estimates. This strategy is analogous to the generalized EM approach of \citet{dempster1977maximum}, which increases, rather than maximizes, the objective at each iteration, and we adopt it throughout our numerical studies.
\end{remark}

\begin{remark}\label{Remark2}
   The key distinction between the two VEM methods---\textsc{JJ-VEM} and \textsc{PG-VEM}---lies in where the approximation enters the inference procedure.
   The \textsc{JJ-VEM} introduces the approximation at the level of the likelihood itself: the Jaakkola--Jordan bound replaces the true extended log-likelihood $\ell_e$ with a surrogate $\ell_e^{\mathrm{JJ}}$, and all subsequent inference---including the EM algorithm and any standard error computation---is based on this surrogate.
   In contrast, \textsc{PG-VEM} employs P\'{o}lya--Gamma data augmentation to rewrite the extended likelihood as an exact augmented likelihood $\ell_e^{\mathrm{PG}}$. The approximation is introduced only at the level of the conditional distribution of the latent variables, through a mean-field assumption.
   Because the P\'{o}lya--Gamma identity is an exact representation, the complete-data log-likelihood $\ell_e^{\mathrm{PG}}$ retains the curvature of the original model. This is the property that makes our approach particularly well suited for uncertainty quantification: the missing information principle can be applied to $\ell_e^{\mathrm{PG}}$ to approximate the observed Fisher information without the systematic bias that arises when it is applied to the surrogate $\ell_e^{\mathrm{JJ}}$.
\end{remark}

\begin{remark}
    \textsc{PG-VEM} and \textsc{PG-VEM-Louis} are not free of approximation error, as both rely on a mean-field variational approximation to the conditional distribution of the latent variables $\boldsymbol{z} = \{\boldsymbol{w}, \boldsymbol{B}\}$ given the observed data $\boldsymbol{y}$. This approximation, by construction, cannot capture the dependence structure of the true conditional distribution, and could therefore, adversely affect both the ideal point estimates and their standard errors. We empirically examine the practical impact of the violation of the mean-field assumption in Appendix~\ref{appendix_D}.
\end{remark}

\subsection{Variational Louis' Method}\label{sub_sec_3.2}
Up to this point, we have focused on the first objective, namely estimating the ideal points. We now turn to the second objective: quantifying uncertainty in these estimates. To this end, we propose a variational Louis' method, to obtain standard errors for the estimated ideal points. We begin by reviewing Louis' method \citep{louis1982finding} in a general setting.

\medskip
\noindent\textbf{Louis' Method.}\\
 Let $\boldsymbol{y}$ denote the observed data, $\boldsymbol{\vartheta}$ the parameter of interest, and $\boldsymbol{z}$ the latent variables. The complete-data log-likelihood is given by
\begin{equation*}
    \ell_c(\boldsymbol{\vartheta}, \boldsymbol{z}; \boldsymbol{y})
    \;\coloneqq\;
    \log p(\boldsymbol{y}, \boldsymbol{z}; \boldsymbol{\vartheta}).
\end{equation*}
Louis' identity \citep{louis1982finding}, which relies on the missing information principle \citep{orchard1972missing}, states that the observed Fisher information matrix
\begin{equation*}
    \mathcal{I}_{\mathrm{obs}}(\boldsymbol{\vartheta})
    \;\coloneqq\;
    -\nabla_{\boldsymbol{\vartheta}}^{2} \log p(\boldsymbol{y}; \boldsymbol{\vartheta}),
\end{equation*}
can be expressed as
\begin{equation}\label{eq:louis}
\begin{split}
    \mathcal{I}_{\mathrm{obs}}(\boldsymbol{\vartheta})
    =
    -\mathbb{E}\!\left[
    \nabla^2_{\boldsymbol{\vartheta}} \ell_c(\boldsymbol{\vartheta}, \boldsymbol{z}; \boldsymbol{y})
    \mid \boldsymbol{y}; \boldsymbol{\vartheta}
    \right]
    -
    \mathrm{Var}\!\left(
    \nabla_{\boldsymbol{\vartheta}} \ell_c(\boldsymbol{\vartheta}, \boldsymbol{z}; \boldsymbol{y})
    \mid \boldsymbol{y}; \boldsymbol{\vartheta}
    \right),
\end{split}
\end{equation}
where the expectation and variance are taken with respect to the conditional distribution $p(\boldsymbol{z}\mid \boldsymbol{y}; \boldsymbol{\vartheta})$. This identity is particularly useful in settings where the marginal likelihood $p(\boldsymbol{y};  \boldsymbol{\vartheta})$ is analytically intractable, but the complete-data log-likelihood $\ell_c(\boldsymbol{\vartheta}, \boldsymbol{z}; \boldsymbol{y})$ admits tractable derivatives.
Louis' method leverages the identity in \eqref{eq:louis} to obtain the standard error of $\hat{\vartheta}_i$ as
\begin{equation*}
    \mathrm{se}(\hat{\vartheta}_i)
    =
    \left\{
    \left(\mathcal{I}_{\mathrm{obs}}^{-1}(\hat{\boldsymbol{\vartheta}})\right)_{(i,i)}
    \right\}^{1/2}.
\end{equation*}

\medskip
\noindent\textbf{Variational Louis' Method.}\\
We now adapt Louis' method to our setting. Let $\sigma_{ij}$ denote the $(i,j)$th entry of $\boldsymbol{\Sigma}_{\tilde{\boldsymbol{\beta}}}$, and parameterize $\boldsymbol{\Sigma}_{\tilde{\boldsymbol{\beta}}}$ by $\boldsymbol{\nu} = \{\sigma_{11}, \sigma_{12}, \sigma_{22}\}$, so that the collection of parameters is redefined as $\boldsymbol{\vartheta} = \{\boldsymbol{\Theta}, \boldsymbol{\nu}\}$. Under this parameterization, the complete-data log-likelihood is $\ell_e^{\mathrm{PG}}(\boldsymbol{\vartheta}, \boldsymbol{z}; \boldsymbol{y})$.
By Louis' identity \eqref{eq:louis}, the observed Fisher information can be expressed as
\begin{equation*}
\begin{split}
    \mathcal{I}_{\mathrm{obs}}(\boldsymbol{\vartheta})
    =
    -\mathbb{E}\!\left[
    \nabla^2_{\boldsymbol{\vartheta}}
    \ell_e^{\mathrm{PG}}(\boldsymbol{\vartheta}, \boldsymbol{z}; \boldsymbol{y})
    \mid \boldsymbol{y}; \boldsymbol{\vartheta}
    \right]
    -
    \mathrm{Var}\!\left(
    \nabla_{\boldsymbol{\vartheta}}
    \ell_e^{\mathrm{PG}}(\boldsymbol{\vartheta}, \boldsymbol{z}; \boldsymbol{y})
    \mid \boldsymbol{y}; \boldsymbol{\vartheta}
    \right).
\end{split}
\end{equation*}
To facilitate the derivation of the standard errors of the ideal points, we partition $\mathcal{I}_{\mathrm{obs}}(\boldsymbol{\vartheta})$ as
\begin{equation*}
    \mathcal{I}_{\mathrm{obs}}(\boldsymbol{\vartheta})=
    \begin{bmatrix}
        \mathcal{I}_{\boldsymbol{\Theta}}(\boldsymbol{\vartheta}) & \mathcal{I}_{\boldsymbol{\Theta}, \boldsymbol{\nu}}(\boldsymbol{\vartheta})\\
        \mathcal{I}_{\boldsymbol{\Theta}, \boldsymbol{\nu}}^\top(\boldsymbol{\vartheta}) & \mathcal{I}_{\boldsymbol{\nu}}(\boldsymbol{\vartheta})
    \end{bmatrix},
\end{equation*}
where
\begin{align*}
    \mathcal{I}_{\boldsymbol{\Theta}}(\boldsymbol{\vartheta})
    &=
    -\mathbb{E}\!\left[
    \nabla^2_{\boldsymbol{\Theta}}
    \ell_e^{\mathrm{PG}}(\boldsymbol{\vartheta}, \boldsymbol{z}; \boldsymbol{y})
    \mid \boldsymbol{y}; \boldsymbol{\vartheta}
    \right]
    -
    \mathrm{Var}\!\left(
    \nabla_{\boldsymbol{\Theta}}
    \ell_e^{\mathrm{PG}}(\boldsymbol{\vartheta}, \boldsymbol{z}; \boldsymbol{y})
    \mid \boldsymbol{y}; \boldsymbol{\vartheta}
    \right),\\[0.25em]
    \mathcal{I}_{\boldsymbol{\nu}}(\boldsymbol{\vartheta})
    &=
    -\mathbb{E}\!\left[
    \nabla^2_{\boldsymbol{\nu}}
    \ell_e^{\mathrm{PG}}(\boldsymbol{\vartheta}, \boldsymbol{z}; \boldsymbol{y})
    \mid \boldsymbol{y}; \boldsymbol{\vartheta}
    \right]
    -
    \mathrm{Var}\!\left(
    \nabla_{\boldsymbol{\nu}}
    \ell_e^{\mathrm{PG}}(\boldsymbol{\vartheta}, \boldsymbol{z}; \boldsymbol{y})
    \mid \boldsymbol{y}; \boldsymbol{\vartheta}
    \right),\\[0.25em]
    \mathcal{I}_{\boldsymbol{\Theta}, \boldsymbol{\nu}}(\boldsymbol{\vartheta})
    &=
    -\mathbb{E}\!\left[
    \nabla_{\boldsymbol{\Theta}}\nabla_{\boldsymbol{\nu}}^\top
    \ell_e^{\mathrm{PG}}(\boldsymbol{\vartheta}, \boldsymbol{z}; \boldsymbol{y})
    \mid \boldsymbol{y}; \boldsymbol{\vartheta}
    \right]
    -
    \mathrm{Cov}\!\big(
    \nabla_{\boldsymbol{\Theta}}
    \ell_e^{\mathrm{PG}}(\boldsymbol{\vartheta}, \boldsymbol{z}; \boldsymbol{y}),
    \nabla_{\boldsymbol{\nu}}
    \ell_e^{\mathrm{PG}}(\boldsymbol{\vartheta}, \boldsymbol{z}; \boldsymbol{y})
    \mid \boldsymbol{y}; \boldsymbol{\vartheta}
    \big).
\end{align*}
Then, the standard error of $\hat{\theta}_i$ is given by
\begin{equation}\label{eq:se_oracle}
    \mathrm{se}(\hat{\theta}_i)
    =
    \left\{
    \left(\mathcal{I}_{\boldsymbol{\Theta}}(\hat{\boldsymbol{\vartheta}})
    -
    \mathcal{I}_{\boldsymbol{\Theta}, \boldsymbol{\nu}}(\hat{\boldsymbol{\vartheta}})
    \mathcal{I}_{\boldsymbol{\nu}}^{-1}(\hat{\boldsymbol{\vartheta}})
    \mathcal{I}_{\boldsymbol{\Theta}, \boldsymbol{\nu}}^\top(\hat{\boldsymbol{\vartheta}})\right)^{-1}
    _{(i,i)}
    \right\}^{1/2}.
\end{equation}

However, the conditional distribution $p(\boldsymbol{z}\mid \boldsymbol{y}; \boldsymbol{\vartheta})$ is not available in closed form. This intractability
motivates the use of a mean-field approximation in the variational E-step of the VEM algorithm. Accordingly, in computing the observed Fisher information, we replace expectations and variances with respect to $p(\boldsymbol{z}\mid \boldsymbol{y}; \boldsymbol{\vartheta})$ by those taken under the variational distribution $q(\boldsymbol{z})$ obtained from the VEM algorithm. This leads to the following variational approximation of $\mathcal{I}_{\mathrm{obs}}(\boldsymbol{\vartheta})$:
\begin{equation*}
\begin{split}
    \tilde{\mathcal{I}}_{\mathrm{obs}}(\boldsymbol{\vartheta})
    &=
    -\mathbb{E}_{q(\boldsymbol{z})}\!\left[
    \nabla^2_{\boldsymbol{\vartheta}}
    \ell_e^{\mathrm{PG}}(\boldsymbol{\vartheta}, \boldsymbol{z}; \boldsymbol{y})
    \right]
    -
    \mathrm{Var}_{q(\boldsymbol{z})}\!\left(
    \nabla_{\boldsymbol{\vartheta}}
    \ell_e^{\mathrm{PG}}(\boldsymbol{\vartheta}, \boldsymbol{z}; \boldsymbol{y})
    \right),
\end{split}
\end{equation*}
with block structure
\begin{equation*}
    \tilde{\mathcal{I}}_{\mathrm{obs}}(\boldsymbol{\vartheta})=
    \begin{bmatrix}
    \tilde{\mathcal{I}}_{\boldsymbol{\Theta}}(\boldsymbol{\vartheta}) & \tilde{\mathcal{I}}_{\boldsymbol{\Theta}, \boldsymbol{\nu}}(\boldsymbol{\vartheta})\\
    \tilde{\mathcal{I}}_{\boldsymbol{\Theta}, \boldsymbol{\nu}}^\top(\boldsymbol{\vartheta}) & \tilde{\mathcal{I}}_{\boldsymbol{\nu}}(\boldsymbol{\vartheta})
    \end{bmatrix}.
\end{equation*}
Under this approximation, the standard error of $\hat{\theta}_i$ is given by
\begin{equation*}
    \mathrm{se}(\hat{\theta}_i)=
    \left\{
    \left(
    \tilde{\mathcal{I}}_{\boldsymbol{\Theta}}(\boldsymbol{\hat{\vartheta}})
    -
    \tilde{\mathcal{I}}_{\boldsymbol{\Theta}, \boldsymbol{\nu}}(\boldsymbol{\hat{\vartheta}})
    \tilde{\mathcal{I}}_{\boldsymbol{\nu}}^{-1}(\boldsymbol{\hat{\vartheta}})
    \tilde{\mathcal{I}}_{\boldsymbol{\Theta}, \boldsymbol{\nu}}^\top(\boldsymbol{\hat{\vartheta}})
    \right)^{-1}
    _{(i,i)}
    \right\}^{1/2},
\end{equation*}
We refer to this procedure as the variational Louis' method. The implementation details of the variational Louis' method are provided in Appendix~\ref{appendix_B}.

\section{Numerical Study}\label{sec_4}
In this section, we evaluate the performance of the proposed method in estimating ideal points and their associated standard errors based on simulation datasets and compare it with existing methods. 

For notational convenience, we denote the proposed method by \textsc{PG-VEM}. As benchmarks, we consider two competing methods: the variational EM method based on the Jaakkola--Jordan approximation (\textsc{JJ-VEM}) and the Bayesian item response theory model (\textsc{BIRT}).
Except for \textsc{BIRT}, for which standard errors are given by the posterior standard deviations of the MCMC samples, both \textsc{PG-VEM} and \textsc{JJ-VEM} provide only ideal point estimates. To obtain standard errors, we additionally employ either the parametric bootstrap (PB) or the variational Louis' method (LOUIS), resulting in four combinations: \textsc{PG-VEM-PB}, \textsc{PG-VEM-Louis}, \textsc{JJ-VEM-PB}, and \textsc{JJ-VEM-Louis}.

\textsc{BIRT} is fitted using the \texttt{ideal} function from the \texttt{pscl} package in \texttt{R}, while all other methods are implemented by our own \texttt{R} code. 
For \textsc{BIRT}, we use a burn-in period of 20{,}000 iterations, followed by 100{,}000 MCMC iterations with a thinning interval of 100; for the prior distributions, we use the default priors provided by the \texttt{ideal} function in the \texttt{pscl} package: independent normal priors with mean zero and precision 1 (i.e., variance 1) for the ideal points, and independent normal priors with mean zero and precision 0.04 (i.e., variance 25) for the bill parameters.
For \textsc{PG-VEM-PB} and \textsc{JJ-VEM-PB}, standard errors are computed by averaging over 100 bootstrap replications.
\subsection{Simulation Scenario~I}
We consider a scenario in which the true ideal points $\theta_i$ are drawn from a bimodal mixture distribution,
\begin{equation*}
    \theta_i \sim 0.5\,N(-2,1) + 0.5\,N(2,1),
    \qquad i = 1,2,\ldots,I,
\end{equation*}
and the bill-specific parameters $\tilde{\boldsymbol{\beta}}_j$ are generated from a bivariate normal distribution,
\begin{equation*}
    \tilde{\boldsymbol{\beta}}_j \sim N_2\!\left(
    \boldsymbol{0},
    \begin{pmatrix}
        2 & 0 \\
        0 & 2
    \end{pmatrix}
    \right),
    \qquad j = 1,2,\ldots,J.
\end{equation*}
Conditional on the ideal points and the bill-specific parameters, the probability that legislator $i$ votes yea on bill $j$ is given by $p_{ij} = \sigma(\alpha_j + \beta_j \theta_i)$, and the observed votes are generated according to
\begin{equation*}
    Y_{ij} \sim \mathrm{Bernoulli}(p_{ij}),
    \qquad i = 1,\ldots,I,\; j = 1,\ldots,J.
\end{equation*}
We set the number of legislators and bills to $I = 400$ and $J = 1{,}000$, respectively.

\begin{figure}[!htb]
    \centering
    \includegraphics[width=1\linewidth]{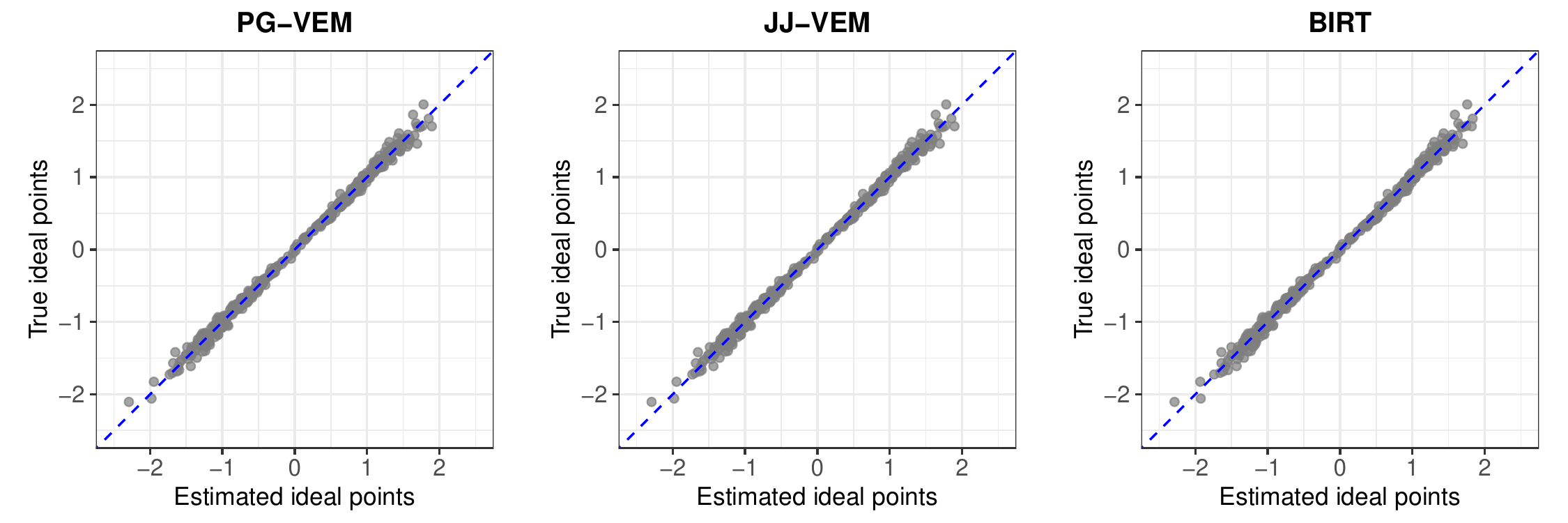}
    \caption{Comparison of the estimated ideal points obtained by \textsc{PG-VEM}, \textsc{JJ-VEM}, and \textsc{BIRT} with the true ideal points in Scenario~I. In each panel, the horizontal axis represents the estimated ideal points, while the vertical axis represents the true ideal points. The blue dashed line indicates the 45-degree line.}
    \label{fig:2}
\end{figure}

To assess how well each method recovers the true ideal points, we compare the estimated ideal points with the corresponding true values. Since ideal points are identifiable only up to an affine transformation of the latent scale, all estimated values are standardized and sign-corrected prior to comparison with the true ideal points. The results are summarized in Figure~\ref{fig:2}. 
All three methods yield nearly identical patterns; most points lie close to the 45-degree line, indicating that the estimated ideal points closely align with the true ideal points; as the true ideal points move further away from zero, the estimation bias tends to increase.

\begin{figure}[!htb]
    \centering
    \includegraphics[width=\linewidth]{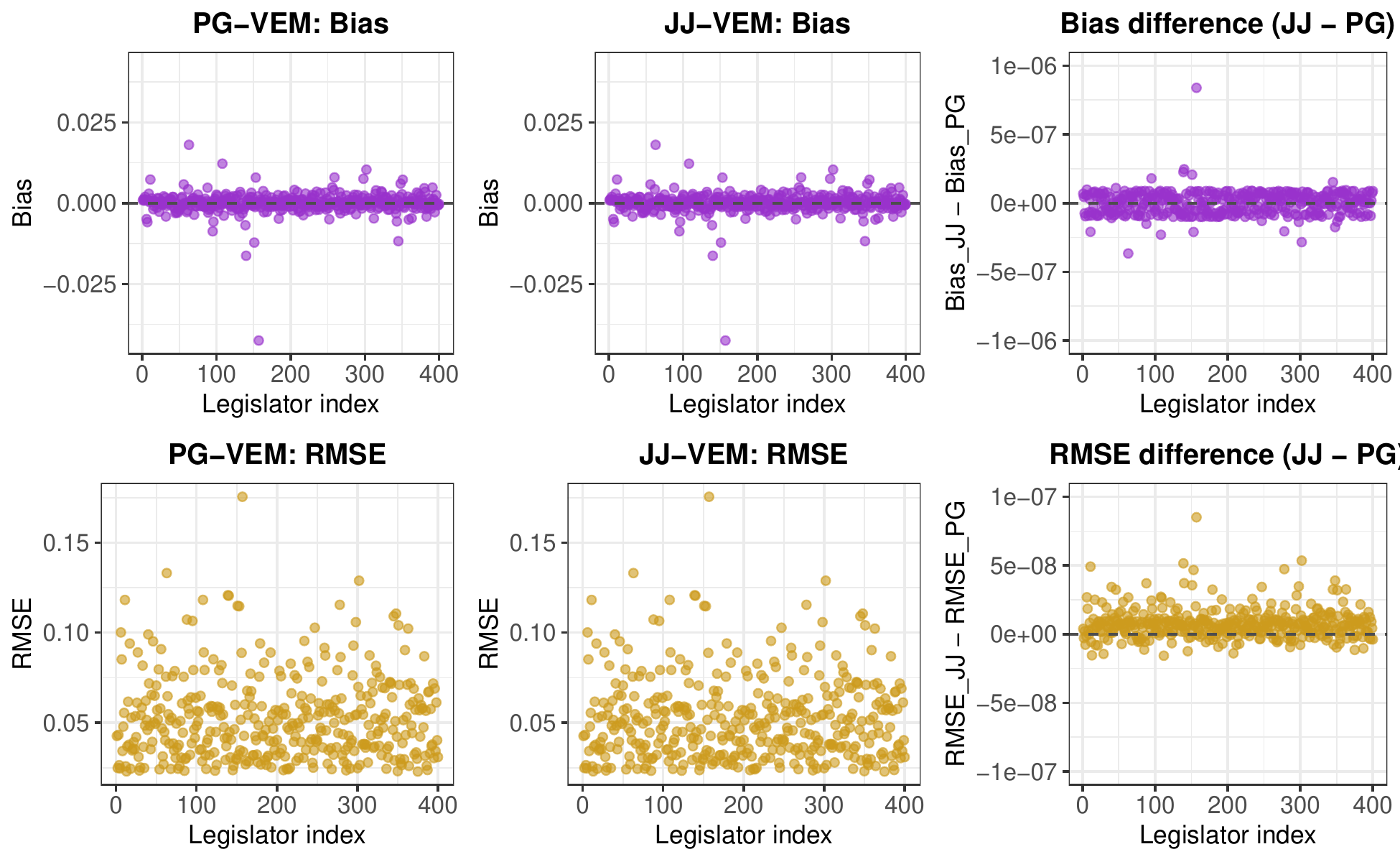}
    \caption{Comparison of the bias and RMSE of \textsc{PG-VEM} and \textsc{JJ-VEM} across $400$ legislators under Simulation Setting~I, based on 1{,}000 replications. Top row: bias by legislator for \textsc{PG-VEM} (left), \textsc{JJ-VEM} (middle), and their difference, \textsc{JJ-VEM} $-$ \textsc{PG-VEM} (right). Bottom row: same as above, but for RMSE.}
    \label{fig:3}
\end{figure}

Next, we compare the bias and RMSE of \textsc{PG-VEM} and \textsc{JJ-VEM} across $400$ legislators, based on 1{,}000 replications. The results are summarized in Figure~\ref{fig:3}.
For bias (top row), the two methods exhibit nearly identical patterns across legislators; the bias remains small overall, with the largest deviation reaching only about $0.04$ in magnitude. The bias difference between the two methods is essentially zero throughout, at most about $10^{-6}$. 
For RMSE (bottom row), the two methods again show similar patterns across legislators. The RMSE difference is positive for most legislators, indicating that \textsc{PG-VEM} exhibits a slightly smaller RMSE than \textsc{JJ-VEM} overall; however, this difference is at most about $10^{-7}$ in magnitude, confirming that the gap between the two methods is negligible in practice. 
Taken together, these results indicate that the two methods are essentially indistinguishable in terms of point estimation accuracy, as measured by bias and RMSE.

\begin{figure}[!htb]
    \centering
    \includegraphics[width=0.6\linewidth]{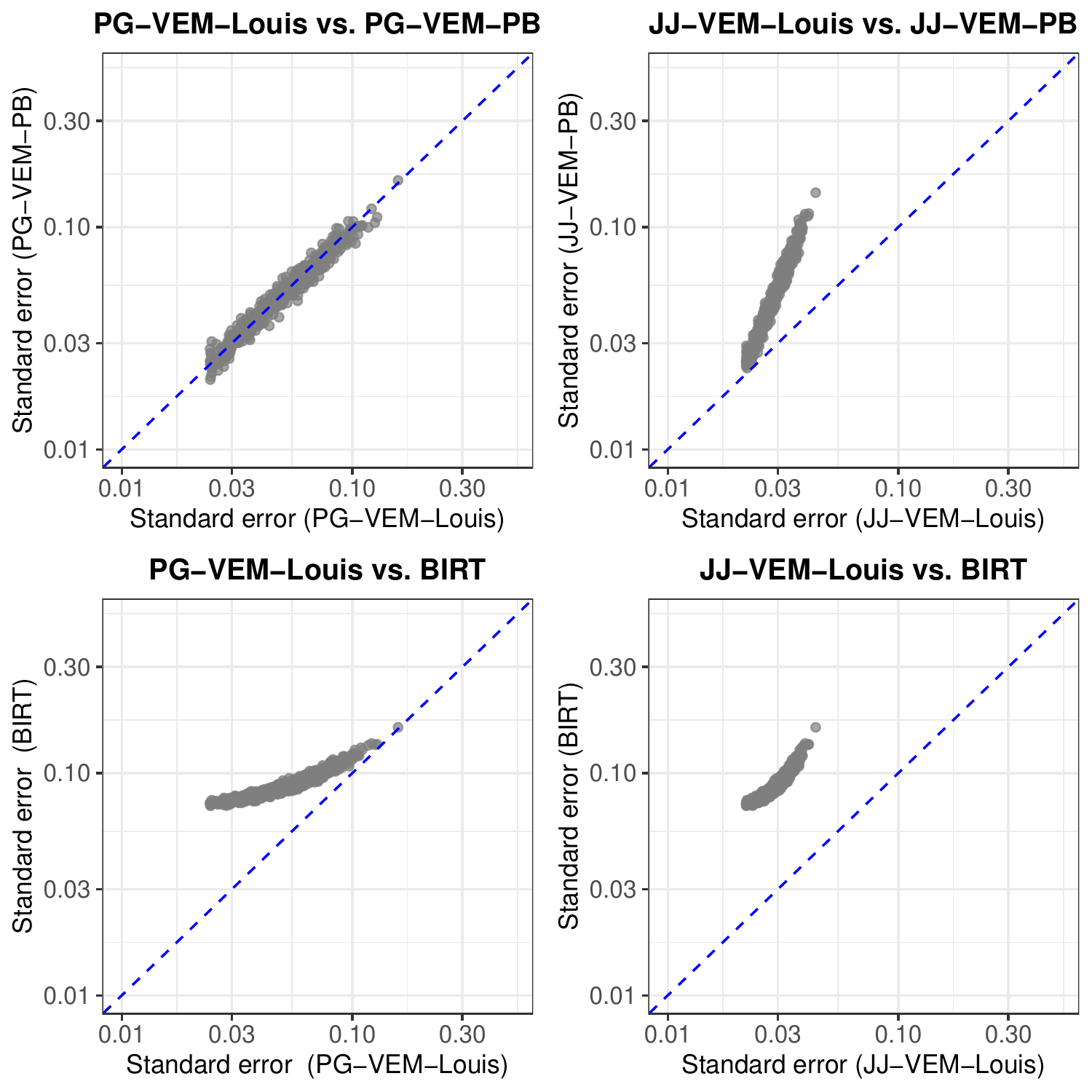}
    \caption{Comparison of standard errors for the estimated ideal points under five methods in Scenario~I: \textsc{PG-VEM-PB}, \textsc{PG-VEM-Louis}, \textsc{JJ-VEM-PB}, \textsc{JJ-VEM-Louis}, and \textsc{BIRT}. In each panel, both axes are shown on a $\log_{10}$ scale, and the blue dashed line indicates the 45-degree line.}
    \label{fig:4}
\end{figure}

Now, we compare the standard errors of the estimated ideal points. The results are summarized in Figure~\ref{fig:4}. Several observations can be made.
First, \textsc{PG-VEM-Louis} yields standard errors nearly identical to those obtained from \textsc{PG-VEM-PB}. Since \textsc{PG-VEM-PB} is regarded as a gold-standard benchmark, this finding justifies the validity of \textsc{PG-VEM-Louis}.
Second, in contrast to \textsc{PG-VEM-Louis}, \textsc{JJ-VEM-Louis} produces standard errors that differ substantially from those obtained via \textsc{JJ-VEM-PB}. As discussed in Remark~\ref{Remark2}, this discrepancy arises because \textsc{JJ-VEM-Louis} applies Louis' method to the variational lower bound $\ell_e^{\mathrm{JJ}}$, whose curvature can differ substantially from that of the true complete-data log-likelihood. In contrast, \textsc{PG-VEM-Louis} applies Louis' method to the P\'{o}lya--Gamma augmented likelihood $\ell_e^{\mathrm{PG}}$, which preserves the curvature of the original model and therefore yields reliable standard errors. This confirms the theoretical advantage of introducing the approximation at the level of the conditional distribution rather than the likelihood itself.
Finally, compared with \textsc{BIRT}, both \textsc{PG-VEM-Louis} and \textsc{JJ-VEM-Louis} tend to produce smaller standard error estimates. This discrepancy is expected and should be interpreted with caution, as it reflects fundamental differences in the underlying modeling frameworks. \textsc{BIRT} adopts a fully Bayesian approach in which both the ideal points and the bill parameters are treated as random variables endowed with prior distributions. As a result, the posterior standard deviations reported by \textsc{BIRT} incorporate both sampling variability and prior uncertainty, and are further influenced by the choice of prior. In contrast, \textsc{PG-VEM} and \textsc{JJ-VEM} treat the ideal points as fixed effects, so that the standard errors reflect only the sampling variability of the maximum likelihood estimator. These differing sources of uncertainty make a direct numerical comparison of the standard errors between the two paradigms inherently difficult to interpret.

\begin{figure}[!htb]
    \centering
    \includegraphics[width=0.7\linewidth]{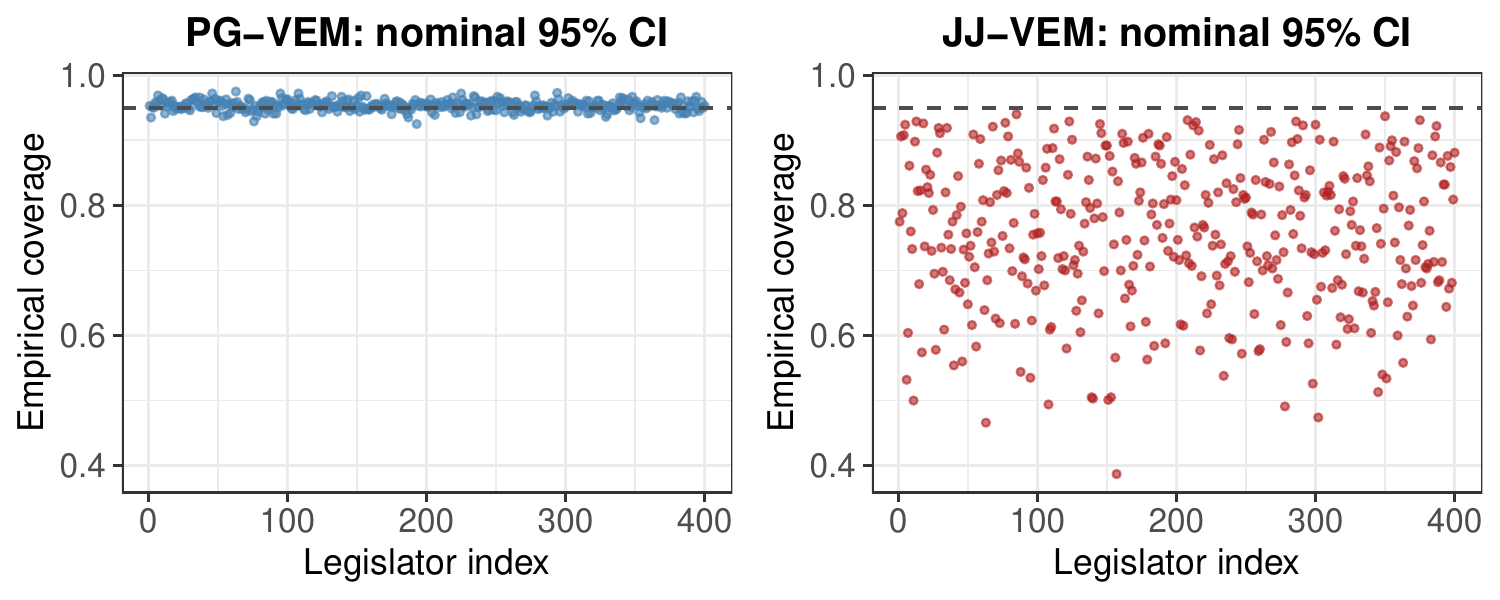}
    \caption{Empirical coverage of confidence interval for the estimated ideal points of $400$ legislators under Simulation Setting~I, based on 1{,}000 replications. Left: \textsc{PG-VEM}. Right: \textsc{JJ-VEM}. The dashed line indicates the nominal 95\% coverage level.}
    \label{fig:5}
\end{figure}

Focusing on \textsc{PG-VEM-Louis} and \textsc{JJ-VEM-Louis}, Figure~\ref{fig:5} shows the empirical coverage of the nominal 95\% confidence interval for each of the $400$ legislators, where the confidence interval for each replication is constructed as $\hat\theta_i \pm 1.96\,\mathrm{se}(\hat\theta_i)$ under the asymptotic normality of the MLE.
For \textsc{PG-VEM} (left panel), the empirical coverage is tightly clustered around the nominal 95\% level across all legislators, indicating that the resulting standard errors are well calibrated. 
In contrast, \textsc{JJ-VEM} (right panel) shows substantial undercoverage for the vast majority of legislators.

\begin{figure}[!htb]
    \centering
    \includegraphics[width=0.8\linewidth]{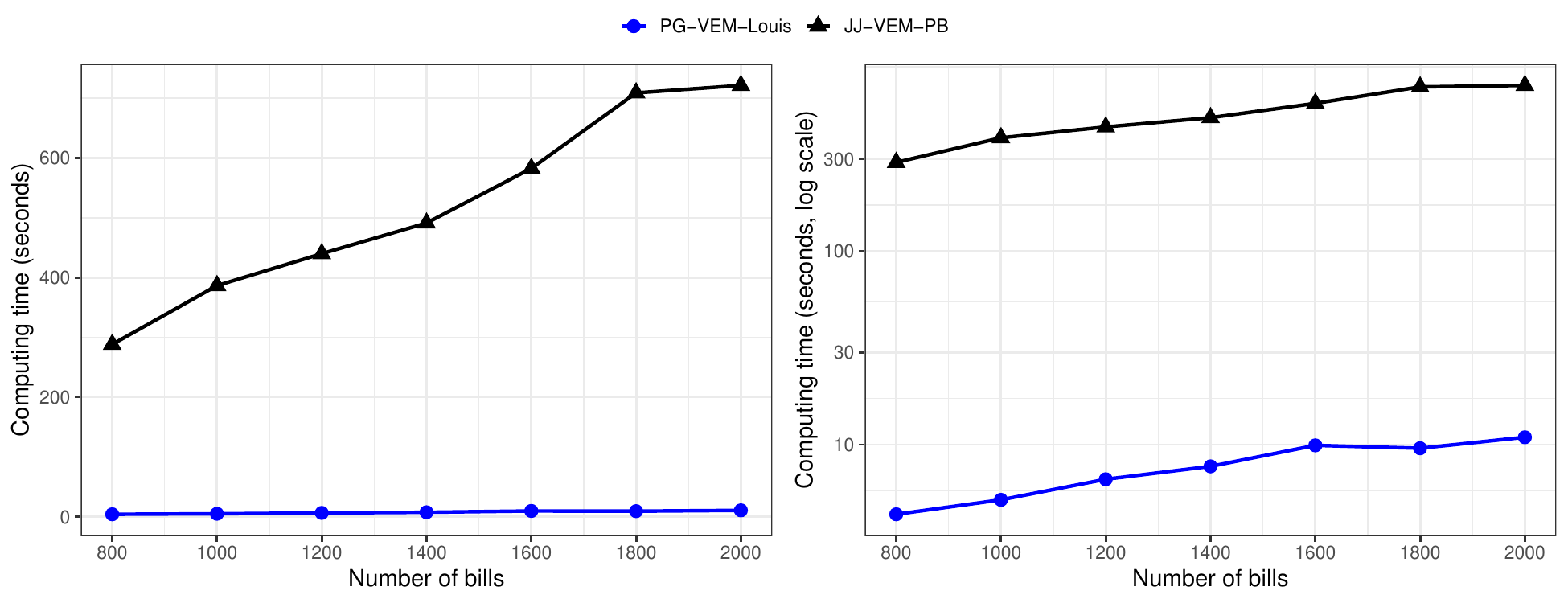}
    \caption{Comparison of computing time for \textsc{PG-VEM-Louis} and \textsc{JJ-VEM-PB} as the number of bills increases ($J \in \{800, 1{,}000, \ldots, 2{,}000\}$), with the number of legislators fixed at $I=400$. The x-axis shows the number of bills, and the y-axis shows the computing time in seconds. The right panel is the same as the left panel, but with the $y$-axis shown on a logarithmic scale, so that the trend in the computing time of \textsc{PG-VEM-Louis} can be clearly seen.}
    \label{fig:6}
\end{figure}

Finally, we compare the computational cost of \textsc{PG-VEM-Louis} and \textsc{JJ-VEM-PB} to illustrate the relative computational efficiency of the proposed method. Since \textsc{BIRT} clearly requires substantially more computation time (even at the smallest setting considered here, $J = 800$, fitting BIRT took nearly an hour), we exclude it from this comparison. Simulated roll-call data are generated as before, with the number of legislators fixed at $I = 400$ and the number of bills varying over $J \in \{800, 1{,}000, \ldots, 2{,}000\}$. The results are summarized in Figure~\ref{fig:6}.
Across all settings, \textsc{PG-VEM-Louis} is computationally more efficient than \textsc{JJ-VEM-PB}, and the gap becomes more pronounced as $J$ increases. Together with the results in Figures~\ref{fig:2} and \ref{fig:4}, these findings indicate that the substantial computational cost of the parametric bootstrap can be avoided by using \textsc{PG-VEM-Louis} without sacrificing the accuracy of ideal point estimates and their standard errors. Consequently, \textsc{PG-VEM-Louis} provides a computationally efficient alternative for estimating ideal points and their standard errors.

\subsection{Simulation Scenario~II}

In this section, we consider a more realistic simulation setting. Following \citet{imai2016fast}, we first estimate the ideal points $\theta_i$ and bill-specific parameters $\tilde{\boldsymbol{\beta}}_j$ by applying \textsc{PG-VEM} to roll-call data from the 112th U.S.\ Congress. Treating these estimates as the ground truth, we generate synthetic roll-call data while preserving the original numbers of legislators and bills (395 legislators and 1{,}455 bills), as well as the original missingness pattern in the roll-call matrix (i.e., an entry in the synthetic data is left missing whenever the corresponding entry in the roll-call data from the 112th U.S.\ Congress was missing, so that the two datasets share the same missingness pattern), thereby ensuring realistic simulation conditions. In roll-call data, missing votes arise for reasons such as a legislator not being in office at the time a bill was voted on, or being absent or abstaining from a vote.

\begin{figure}[!htb]
    \centering
    \includegraphics[width=\linewidth]{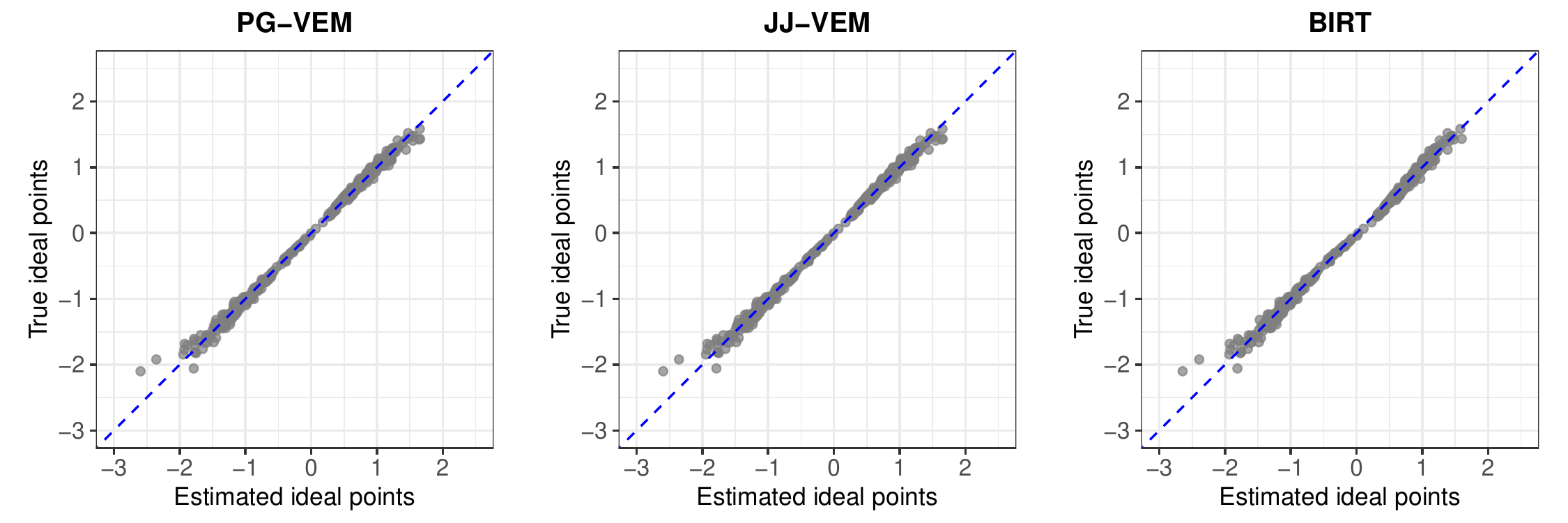}
    \caption{Comparison of the estimated ideal points obtained by \textsc{PG-VEM}, \textsc{JJ-VEM}, and \textsc{BIRT} with the true ideal points in Scenario~II. In each panel, the horizontal axis represents the estimated ideal points, while the vertical axis represents the true ideal points. The blue dashed line indicates the 45-degree line.}
    \label{fig:7}
\end{figure}

Figure~\ref{fig:7} compares the estimated ideal points with the corresponding true values. All methods produce very similar patterns and exhibit good recovery of the true ideal points.

\begin{figure}[!htb]
    \centering
    \includegraphics[width=0.6\linewidth]{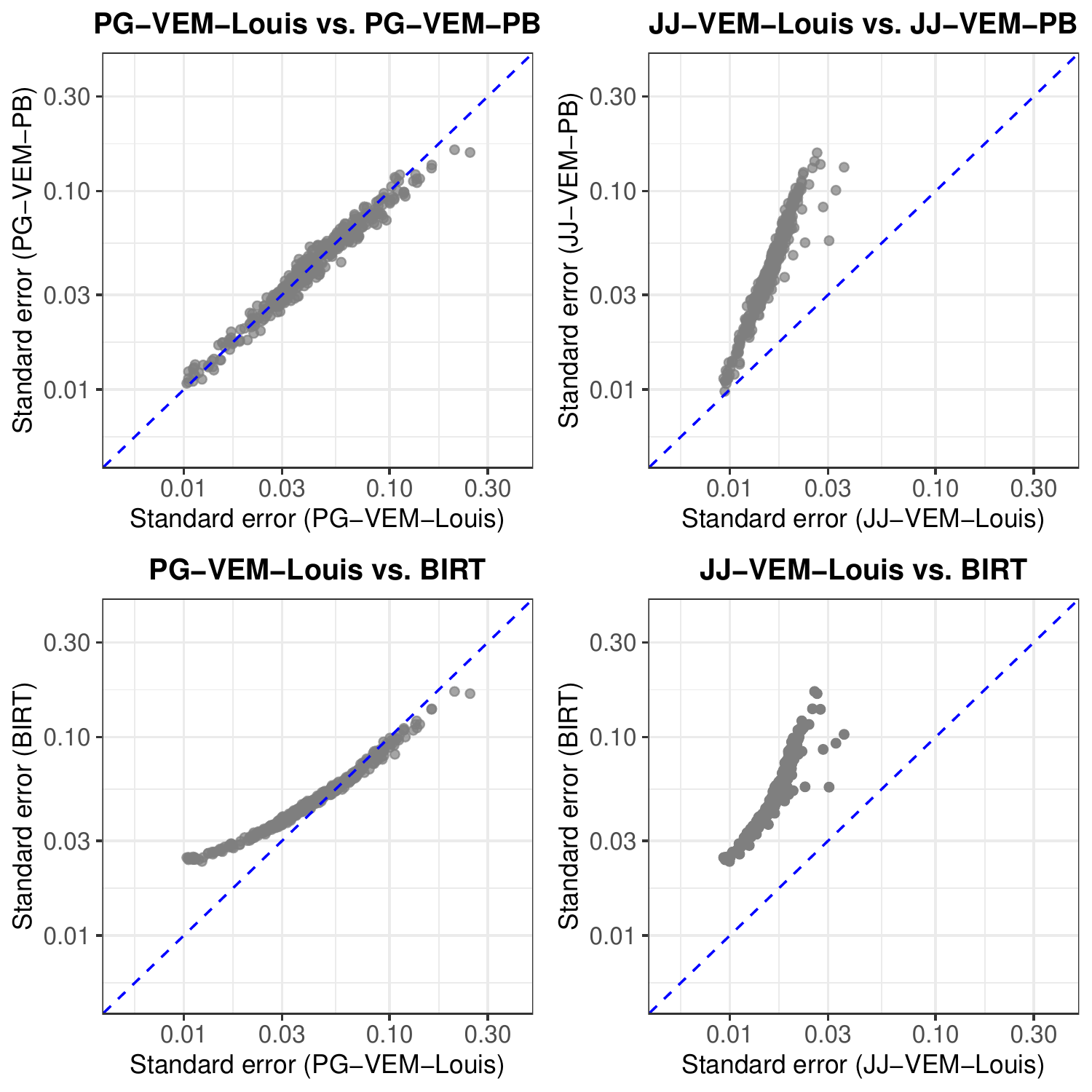}
    \caption{Comparison of standard errors for the estimated ideal points under five methods in Scenario~II: \textsc{PG-VEM-PB}, \textsc{PG-VEM-Louis}, \textsc{JJ-VEM-PB}, \textsc{JJ-VEM-Louis}, and \textsc{BIRT}. In each panel, both axes are shown on a $\log_{10}$ scale, and the blue dashed line indicates the 45-degree line.}
    \label{fig:8}
\end{figure}

Figure~\ref{fig:8} compares the estimated standard errors. Overall, patterns similar to those observed in Scenario~I emerge: the standard errors from \textsc{PG-VEM-Louis} closely match those from \textsc{PG-VEM-PB}. In contrast, \textsc{JJ-VEM-Louis} continues to exhibit substantial deviations from the corresponding parametric bootstrap estimates obtained from \textsc{JJ-VEM-PB}. This further corroborates the finding from Scenario~I that applying Louis' method to the variational lower bound $\ell_e^{\mathrm{JJ}}$, rather than to an exact augmented likelihood, leads to unreliable standard errors. Compared with \textsc{BIRT}, both \textsc{PG-VEM-Louis} and \textsc{JJ-VEM-Louis} tend to produce smaller standard error estimates, consistent with the discussion of differing modeling assumptions in Scenario~I.

\begin{figure}[!htb]
    \centering
    \includegraphics[width=\linewidth]{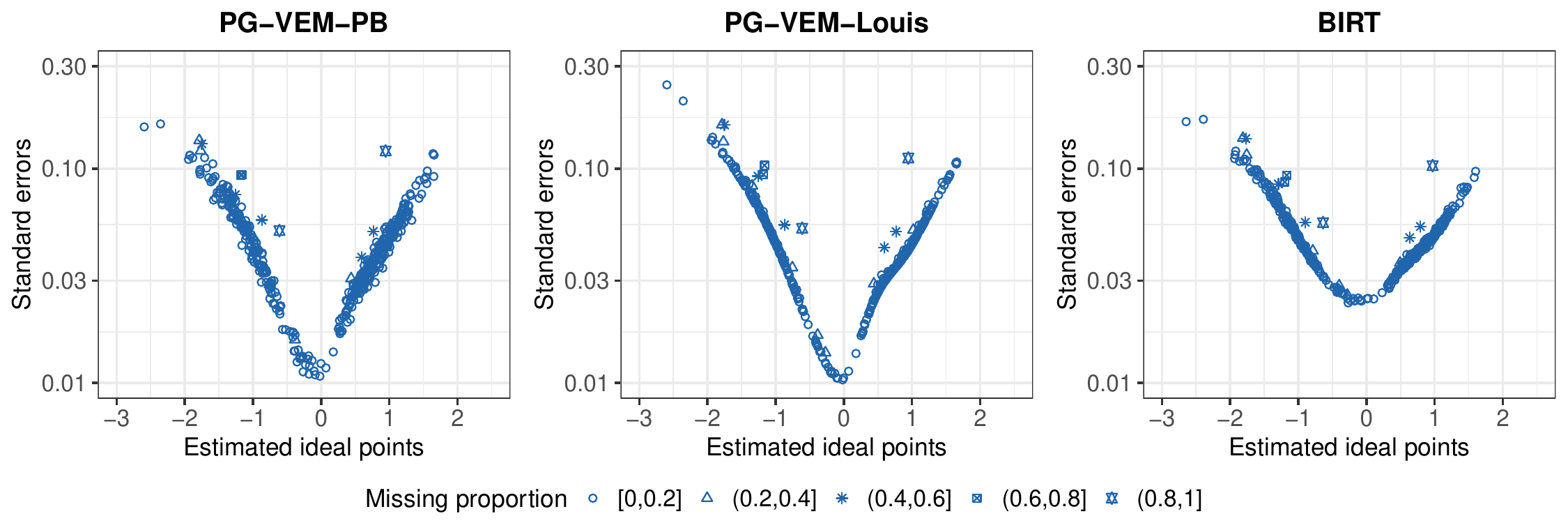}
    \caption{Scatter plots of estimated ideal points (horizontal axis) versus estimated standard errors (vertical axis, on the $\log_{10}$ scale), with point shapes indicating the missing rate. From left to right, the three panels correspond to the results from \textsc{PG-VEM-PB}, \textsc{PG-VEM-Louis}, and \textsc{BIRT}.}
    \label{fig:9}
\end{figure}

Figure~\ref{fig:9} shows scatter plots of the estimated ideal points (horizontal axis) versus the estimated standard errors (vertical axis), with the missing rate indicated by the point shapes. We make the following observations. First, the estimated standard errors tend to increase as the estimated ideal points deviate further from zero. Second, even for similar estimated ideal points, the estimated standard errors increase when the corresponding observations have higher missing rates. Third, \textsc{PG-VEM-PB} and \textsc{PG-VEM-Louis} exhibit very similar patterns, providing further empirical support for the validity of \textsc{PG-VEM-Louis}. Finally, compared with \textsc{BIRT}, \textsc{PG-VEM-Louis} tends to produce smaller standard error estimates, particularly when the estimated ideal points are close to zero, a pattern consistent with Figure~\ref{fig:8}.

\section{Real Data Example: the 113th U.S. Congress}\label{sec_5}
\begin{figure}[!htb]
    \centering
    \includegraphics[width=\linewidth]{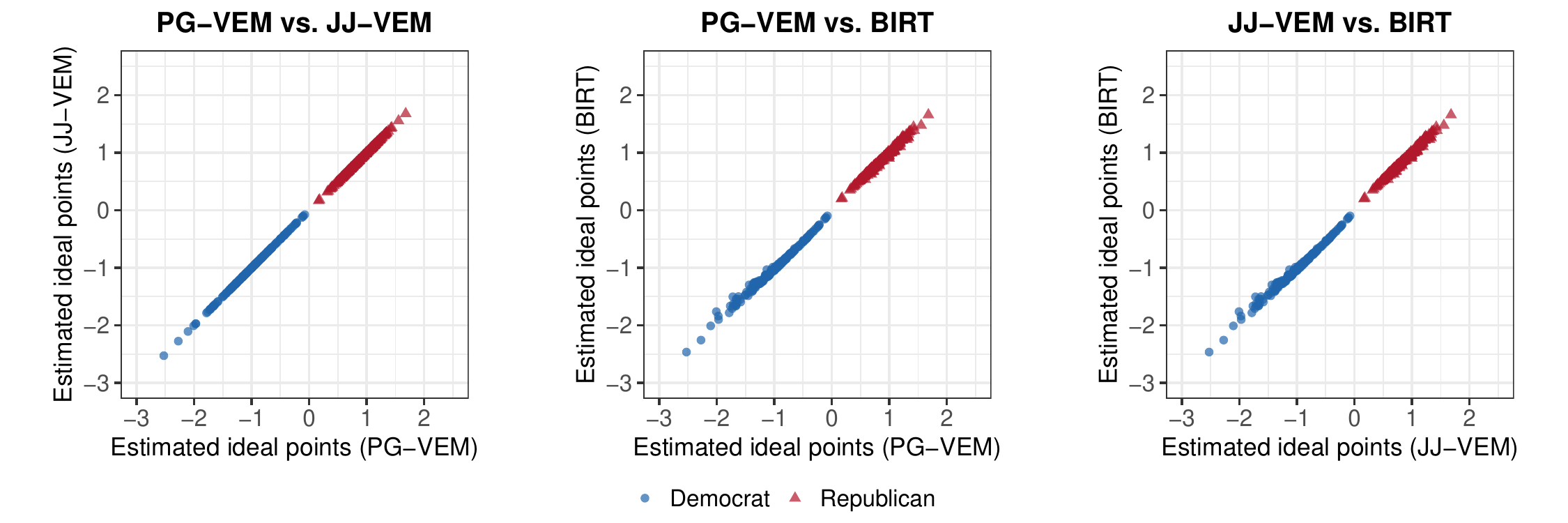}
    \caption{The 113th U.S. Congress: comparison of the estimated ideal points obtained by \textsc{PG-VEM}, \textsc{JJ-VEM}, and \textsc{BIRT}. Blue circles denote Democrats, and red triangles denote Republicans.}
    \label{fig:10}
\end{figure}

\begin{figure}[!htb]
    \centering
    \includegraphics[width=0.6\linewidth]{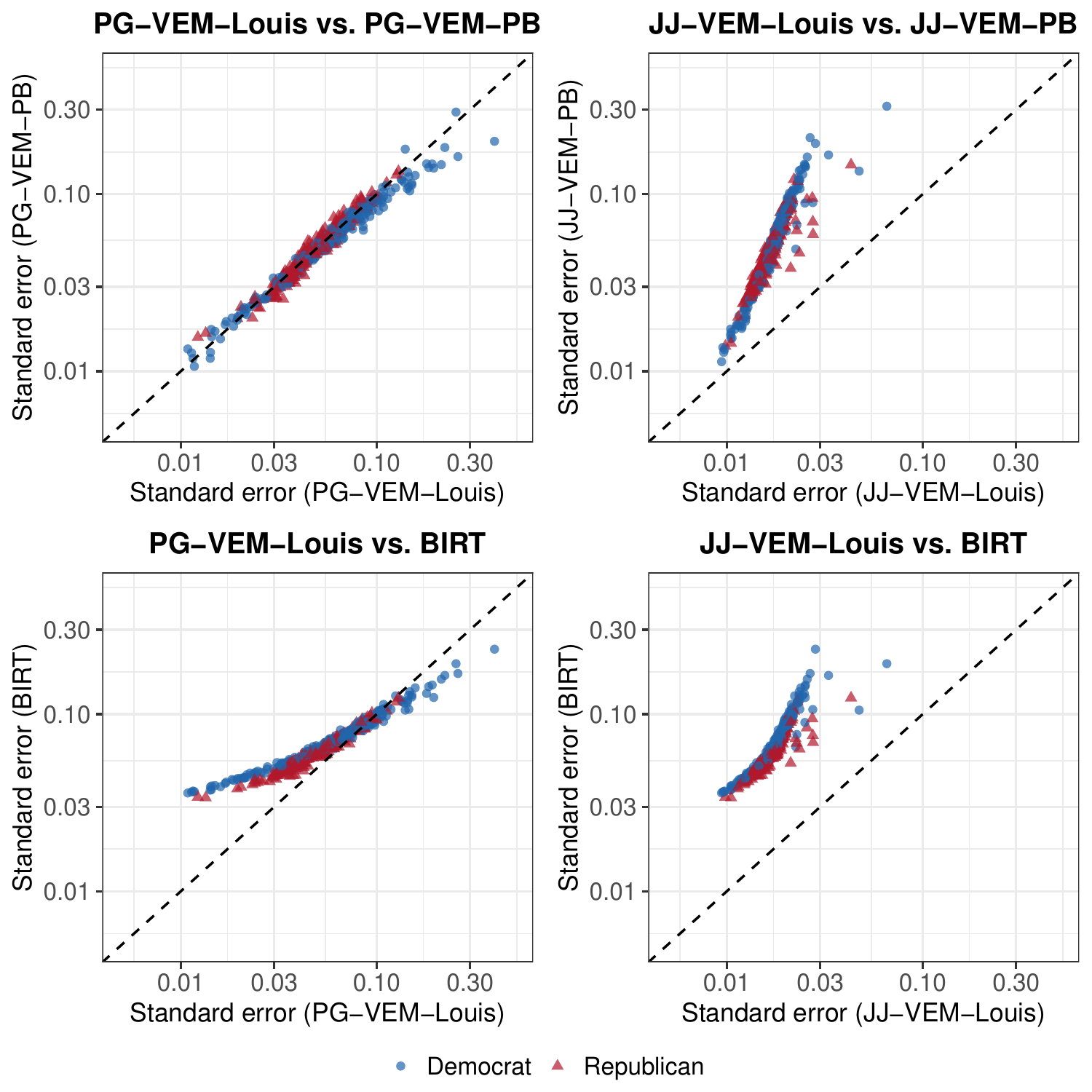}
    \caption{The 113th U.S. Congress: comparison of standard errors for the estimated ideal points under five methods---\textsc{PG-VEM-PB}, \textsc{PG-VEM-Louis}, \textsc{JJ-VEM-PB}, \textsc{JJ-VEM-Louis}, and \textsc{BIRT}. In each panel, both axes are on a $\log_{10}$ scale, and the black dashed line represents the 45-degree line. Blue circles indicate Democrats, and red triangles indicate Republicans.}
    \label{fig:11}
\end{figure}

\begin{figure}[!htb]
    \centering
    \includegraphics[width=\linewidth]{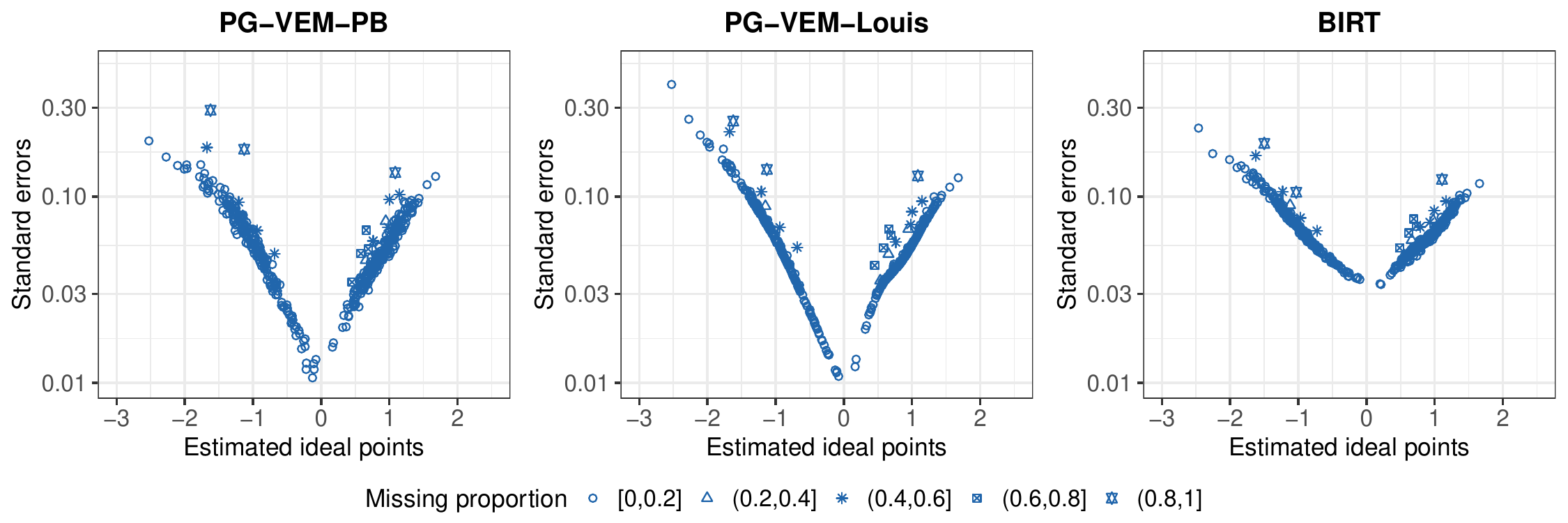}
    \caption{The 113th U.S. Congress: scatter plots of estimated ideal points (horizontal axis) versus estimated standard errors (vertical axis, on the $\log_{10}$ scale), with point shapes indicating the missing rate. From left to right, the three panels correspond to the results from \textsc{PG-VEM-PB}, \textsc{PG-VEM-Louis}, and \textsc{BIRT}.}
    \label{fig:12}
\end{figure}

In this section, we apply \textsc{PG-VEM}, \textsc{JJ-VEM}, and \textsc{BIRT} to roll-call data from the 113th U.S.\ Congress. The results are summarized in Figures~\ref{fig:10}--\ref{fig:12}. 
An additional real-data analysis based on the 118th U.S.\ Congress, which exhibits patterns consistent with those observed for the 113th U.S.\ Congress, is provided in Appendix~\ref{appendix_C}.

Figure~\ref{fig:10} compares the estimated ideal points across different methods. Consistent with the simulation results, all methods produce highly similar ideal point estimates.

Figure~\ref{fig:11} compares the estimated standard errors. The standard errors from \textsc{PG-VEM-Louis} closely match those obtained from \textsc{PG-VEM-PB}, whereas \textsc{JJ-VEM-Louis} exhibits substantial deviations from \textsc{JJ-VEM-PB}. These patterns are consistent with the simulation findings, corroborating the theoretical argument that preserving the curvature of the complete-data log-likelihood is essential for reliable standard error estimation via Louis' method (see Remark~\ref{Remark2}). Compared with \textsc{BIRT}, the variational Louis' methods tend to produce smaller standard error estimates, reflecting the distinct treatment of the ideal points as fixed effects in our framework versus random variables in the Bayesian approach.

Figure~\ref{fig:12} displays scatter plots of the estimated ideal points against their corresponding standard errors, with point shapes indicating the missing rate. The same qualitative patterns observed in Figure~\ref{fig:9} persist in the real data: standard errors increase as the estimated ideal points move away from zero, and for similar ideal point estimates, higher missing rates are associated with larger standard errors. Moreover, the close agreement between \textsc{PG-VEM-PB} and \textsc{PG-VEM-Louis} provides further empirical support for the validity of \textsc{PG-VEM-Louis}. Compared with \textsc{BIRT}, \textsc{PG-VEM-Louis} tends to underestimate the standard errors when the estimated ideal points are close to zero.

\section{Conclusion}\label{sec_6}
We propose a new method for estimating legislators' ideal points and their associated standard errors. The key idea of the proposed approach is to leverage the P\'{o}lya--Gamma identity to circumvent the intractable integral arising in the marginal likelihood. Based on this representation, we derive a VEM algorithm, referred to as \textsc{PG-VEM}, for estimating the ideal points. We further develop a variational Louis' method that utilizes the variational approximation of the conditional distribution of latent variables obtained from \textsc{PG-VEM} to compute the associated standard errors. A key advantage of the proposed approach over existing variational methods is that the P\'{o}lya--Gamma identity provides an exact augmented likelihood, so that the approximation enters only through the mean-field assumption on the conditional distribution of latent variables rather than through the likelihood itself. This preserves the curvature of the complete-data log-likelihood and enables reliable application of the missing information principle for standard error estimation. Extensive numerical experiments and real data analysis consistently show that the proposed method produces ideal point estimates comparable to those obtained from existing approaches and yields standard errors close to those obtained from the parametric bootstrap, which is often regarded as a gold standard. The computational cost of the proposed method is substantially lower than that of MCMC-based methods or parametric bootstrap procedures. These results indicate that the proposed method provides a computationally efficient alternative for estimating ideal points and their standard errors.

An important direction for future research is to develop asymptotic theory, such as consistency, for the ideal point estimates and associated standard errors obtained from \textsc{PG-VEM} and \textsc{PG-VEM-Louis}. 
In the literature, theoretical justification for variational approximations remains very limited and is currently available only for simple models---e.g., Bayesian linear regression \citep{you2014variational}, Poisson mixed-effects models \citep{hall2011theory}, and stochastic block models \citep{celisse2012consistency}---as reviewed in Section~5.2 of \citet{blei2017variational}.
To the best of our knowledge, no such theoretical guarantees currently exist for the IRT model; one reason for this gap is its complex dependence structure among the observations.

Before concluding, we address a question that naturally arises: whether the variational EM algorithm and the variational Louis' method proposed in this paper can also be applied to a probit link-based IRT model. This is indeed possible. Specifically, using the Albert--Chib data augmentation \citep{albert1993bayesian} in place of the P\'{o}lya--Gamma augmentation, we can develop a variational EM algorithm and a variational Louis' method that closely parallel those developed in this paper for the logistic link. However, the logistic distribution has heavier tails than the normal distribution, which may make ideal point estimates somewhat more robust to outlying voting behavior than those based on the probit link; for this reason, unless there is a specific reason to prefer the probit link, we recommend using the logistic-link-based variational EM algorithm and variational Louis' method in practical applications.

\FloatBarrier
\section*{Acknowledgement}
The authors declare that there are no conflicts of interest.

\section*{Data availability}
The roll call voting data used in this study are publicly available from the Voteview database (\url{https://voteview.com/}).

\section*{Funding}
This work was supported by National Research Foundation of Korea (NRF) (No. RS-2025-00520739).
\bibliographystyle{apalike}
\bibliography{ref}

@article{imai2016fast,
  title={Fast estimation of ideal points with massive data},
  author={Imai, Kosuke and Lo, James and Olmsted, Jonathan},
  journal={American Political Science Review},
  volume={110},
  number={4},
  pages={631--656},
  year={2016}
}

@article{cho2021gaussian,
  title={Gaussian variational estimation for multidimensional item response theory},
  author={Cho, April E and Wang, Chun and Zhang, Xue and Xu, Gongjun},
  journal={British Journal of Mathematical and Statistical Psychology},
  volume={74},
  pages={52--85},
  year={2021}
}

@inproceedings{jaakkola1997variational,
  title={A variational approach to Bayesian logistic regression models and their extensions},
  author={Jaakkola, Tommi S and Jordan, Michael I},
  booktitle={Sixth International Workshop on Artificial Intelligence and Statistics},
  pages={283--294},
  year={1997}
}

@book{bishop2006pattern,
  title     = {Pattern Recognition and Machine Learning},
  author    = {Bishop, Christopher M.},
  series    = {Information Science and Statistics},
  publisher = {Springer},
  address   = {New York, NY},
  year      = {2006}
}

@article{blei2017variational,
  title={Variational inference: A review for statisticians},
  author={Blei, David M and Kucukelbir, Alp and McAuliffe, Jon D},
  journal={Journal of the American Statistical Association},
  volume={112},
  number={518},
  pages={859--877},
  year={2017}
}

@article{clinton2004statistical,
  title={The statistical analysis of roll call data},
  author={Clinton, Joshua and Jackman, Simon and Rivers, Douglas},
  journal={American Political Science Review},
  volume={98},
  number={2},
  pages={355--370},
  year={2004}
}

@article{poole1985spatial,
  title={A spatial model for legislative roll call analysis},
  author={Poole, Keith T and Rosenthal, Howard},
  journal={American Journal of Political Science},
  volume = {29},
  number = {2},
  pages={357--384},
  year={1985}
}

@article{xiao2024note,
  title={A Note on Standard Errors for Multidimensional Two-Parameter Logistic Models Using Gaussian Variational Estimation},
  author={Xiao, Jiaying and Wang, Chun and Xu, Gongjun},
  journal={Applied Psychological Measurement},
  volume={48},
  number={6},
  pages={276--294},
  year={2024}
}

@article{polson2013bayesian,
  title={Bayesian inference for logistic models using {P}{\'o}lya--{G}amma latent variables},
  author={Polson, Nicholas G and Scott, James G and Windle, Jesse},
  journal={Journal of the American Statistical Association},
  volume={108},
  number={504},
  pages={1339--1349},
  year={2013}
}

@article{carroll2009measuring,
  title={Measuring bias and uncertainty in {DW-NOMINATE} ideal point estimates via the parametric bootstrap},
  author={Carroll, Royce and Lewis, Jeffrey B and Lo, James and Poole, Keith T and Rosenthal, Howard},
  journal={Political Analysis},
  volume={17},
  number={3},
  pages={261--275},
  year={2009}
}

@article{birnbaum1968some,
  title={Some latent trait models and their use in inferring an examinee's ability},
  author={Birnbaum, Allan},
  journal={Statistical Theories of Mental Test Scores},
  year={1968},
  publisher={Addison-Wesley}
}

@article{berger1988likelihood,
 author = {James O. Berger and Robert L. Wolpert and M. J. Bayarri and M. H. DeGroot and Bruce M. Hill and David A. Lane and Lucien LeCam},
 title = {The Likelihood Principle},
 journal = {Lecture Notes-Monograph Series},
 volume = {6},
 pages = {iii--199},
 year = {1988}
}

@article{louis1982finding,
  title={Finding the observed information matrix when using the EM algorithm},
  author={Louis, Thomas A},
  journal={Journal of the Royal Statistical Society Series B: Statistical Methodology},
  volume={44},
  number={2},
  pages={226--233},
  year={1982}
}

@inproceedings{orchard1972missing,
  title={A missing information principle: theory and applications},
  author={Orchard, Terence and Woodbury, Max A},
  booktitle={Proceedings of the Sixth Berkeley Symposium on Mathematical Statistics and Probability, Volume 1: Theory of Statistics},
  volume={6},
  pages={697--716},
  year={1972}
}

@article{jordan1999introduction,
  title={An introduction to variational methods for graphical models},
  author={Jordan, Michael I and Ghahramani, Zoubin and Jaakkola, Tommi S and Saul, Lawrence K},
  journal={Machine Learning},
  volume={37},
  number={2},
  pages={183--233},
  year={1999}
}

@article{kan2008moments,
  title={From moments of sum to moments of product},
  author={Kan, Raymond},
  journal={Journal of Multivariate Analysis},
  volume={99},
  number={3},
  pages={542--554},
  year={2008}
}

@article{lewis2004measuring,
  title={Measuring bias and uncertainty in ideal point estimates via the parametric bootstrap},
  author={Lewis, Jeffrey B and Poole, Keith T},
  journal={Political Analysis},
  volume={12},
  number={2},
  pages={105--127},
  year={2004}
}

@article{poole2001geometry,
  title={The geometry of multidimensional quadratic utility in models of parliamentary roll call voting},
  author={Poole, Keith T},
  journal={Political Analysis},
  volume={9},
  number={3},
  pages={211--226},
  year={2001}
}

@article{carroll2013structure,
  title={The structure of utility in spatial models of voting},
  author={Carroll, Royce and Lewis, Jeffrey B and Lo, James and Poole, Keith T and Rosenthal, Howard},
  journal={American Journal of Political Science},
  volume={57},
  number={4},
  pages={1008--1028},
  year={2013}
}

@article{shin2025,
  title={$\ell_1$-based Bayesian Ideal Point Model for Multidimensional Politics},
  author={Shin, Sooahn and Lim, Johan and Park, Jong Hee},
  journal={Journal of the American Statistical Association},
  volume={120},
  number={550},
  pages={631--644},
  year={2025}
}

@book{enelow1984spatial,
  title={The Spatial Theory of Voting: An Introduction},
  author={Enelow, James M and Hinich, Melvin J},
  year={1984},
  address={New York},
  publisher={Cambridge University Press}
}

@article{lee1996hierarchical,
  title={Hierarchical generalized linear models},
  author={Lee, Youngjo and Nelder, John A},
  journal={Journal of the Royal Statistical Society Series B: Statistical Methodology},
  volume={58},
  number={4},
  pages={619--656},
  year={1996}
}

@article{jin2024standard,
  title={Standard error estimates in hierarchical generalized linear models},
  author={Jin, Shaobo and Lee, Youngjo},
  journal={Computational Statistics \& Data Analysis},
  volume={189},
  pages={107852},
  year={2024}
}

@article{lee2001hierarchical,
  title={Hierarchical generalised linear models: a synthesis of generalised linear models, random-effect models and structured dispersions},
  author={Lee, Youngjo and Nelder, John A},
  journal={Biometrika},
  volume={88},
  number={4},
  pages={987--1006},
  year={2001}
}

@article{dempster1977maximum,
  title={Maximum likelihood from incomplete data via the EM algorithm},
  author={Dempster, Arthur P and Laird, Nan M and Rubin, Donald B},
  journal={Journal of the royal statistical society: series B (methodological)},
  volume={39},
  number={1},
  pages={1--22},
  year={1977},
  publisher={Wiley Online Library}
}

@article{albert1993bayesian,
  title={Bayesian analysis of binary and polychotomous response data},
  author={Albert, James H and Chib, Siddhartha},
  journal={Journal of the American statistical Association},
  volume={88},
  number={422},
  pages={669--679},
  year={1993},
  publisher={Taylor \& Francis}
}

@article{you2014variational,
  title={On variational Bayes estimation and variational information criteria for linear regression models},
  author={You, Chong and Ormerod, John T and Mueller, Samuel},
  journal={Australian \& New Zealand Journal of Statistics},
  volume={56},
  number={1},
  pages={73--87},
  year={2014},
  publisher={Wiley Online Library}
}

@article{hall2011theory,
  title={Theory of Gaussian variational approximation for a Poisson mixed model},
  author={Hall, Peter and Ormerod, John T and Wand, Matt P},
  journal={Statistica Sinica},
  volume = {21},
  number = {1},
  pages={369--389},
  year={2011},
  publisher={JSTOR}
}

@article{celisse2012consistency,
  title={Consistency of maximum-likelihood and variational estimators in the stochastic block model},
  author={Celisse, Alain and Daudin, Jean-Jacques and Pierre, Laurent},
  journal = {Electronic Journal of Statistics},
  volume = {6},
  pages = {1847--1899},
  year={2012}
}

\appendix
\numberwithin{equation}{section}
\numberwithin{table}{section}
\numberwithin{figure}{section}

\numberwithin{theorem}{section}
\numberwithin{lemma}{section}
\numberwithin{proposition}{section}
\numberwithin{corollary}{section}

\section{VEM Algorithm with Jaakkola--Jordan Approximation}\label{appendix_A}

\noindent\textbf{E-step.}\\
Let $\boldsymbol{\theta}^{(t)} = \{\theta_i^{(t)}\}$, $\boldsymbol{\xi}^{(t)} = \{\xi_{ij}^{(t)}\}$ and $\boldsymbol{\Sigma}_{\tilde{\boldsymbol{\beta}}}^{(t)}$ denote the estimates of the ideal points, the variational parameters and covariance matrix $\boldsymbol{\Sigma}_{\tilde{\boldsymbol{\beta}}}$ at iteration $t$. We first evaluate the conditional distribution of the latent random effects $\tilde{\boldsymbol{\beta}}_j$ given the observed data and the current parameter values, and then use this distribution to compute the corresponding expected complete log-likelihood (the Q-function).
With some algebra, the conditional distribution of $\tilde{\boldsymbol{\beta}}_j$ takes the form
\begin{equation*}
    \tilde{\boldsymbol{\beta}}_j 
    \mid 
    \boldsymbol{y}, \boldsymbol{\theta}^{(t)}, \boldsymbol{\xi}^{(t)}
    \sim 
    N\!\left(
        \tilde{\boldsymbol{\mu}}_j^{(t)},
        \;\;
        \tilde{\boldsymbol{\Sigma}}_j^{(t)}
    \right),
\end{equation*}
where
\begin{equation*}
\begin{split}
    \tilde{\boldsymbol{\mu}}_j^{(t)} 
    = 
    \tilde{\boldsymbol{\Sigma}}_j^{(t)}
    \sum_{i=1}^I (y_{ij}-\frac{1}{2})\,\tilde{\boldsymbol{\theta}}_i^{(t)},
    \quad
    \tilde{\boldsymbol{\Sigma}}_j^{(t)}
    =
    \left(
    \left(\boldsymbol{\Sigma}_{\tilde{\boldsymbol{\beta}}}^{(t)}\right)^{-1}
    - 
    2\sum_{i=1}^I 
    \lambda(\xi_{ij}^{(t)})\,
    \tilde{\boldsymbol{\theta}}_i^{(t)}
    \tilde{\boldsymbol{\theta}}_i^{(t)\top}
    \right)^{-1}.
\end{split}
\end{equation*}
Based on this conditional distribution, the Q-function is defined as
\begin{align*}
    &Q\!\left(
    \boldsymbol{\theta}, \boldsymbol{\Sigma}_{\tilde{\boldsymbol{\beta}}}, \boldsymbol{\xi}
    \,\middle|\,
    \boldsymbol{\theta}^{(t)}, \boldsymbol{\Sigma}_{\tilde{\boldsymbol{\beta}}}^{(t)}, \boldsymbol{\xi}^{(t)}
    \right)\\
    &\quad\coloneqq
    \mathbb{E}\!\left[
    \tilde{\ell}_e\!\left(
    \boldsymbol{\theta}, \boldsymbol{\Sigma}_{\tilde{\boldsymbol{\beta}}},\boldsymbol{\xi}, \boldsymbol{B}; \boldsymbol{y}
    \right)
    \,\middle|\,
    \boldsymbol{y},
    \boldsymbol{\theta}^{(t)},
    \boldsymbol{\Sigma}_{\tilde{\boldsymbol{\beta}}}^{(t)},
    \boldsymbol{\xi}^{(t)}
    \right] \\[0.5em]
    &\quad=
    \sum_{i=1}^I \sum_{j=1}^J
    \Bigg[
    y_{ij}\tilde{\boldsymbol{\mu}}_j^{(t)\top}\tilde{\boldsymbol{\theta}}_i
    + \log \sigma(\xi_{ij})
    - \frac{\tilde{\boldsymbol{\mu}}_j^{(t)\top}\tilde{\boldsymbol{\theta}}_i + \xi_{ij}}{2}
    + \lambda(\xi_{ij})
    \left\{
    \tilde{\boldsymbol{\theta}}_i^\top
    \bigl(\tilde{\boldsymbol{\Sigma}}_j^{(t)}
    + \tilde{\boldsymbol{\mu}}_j^{(t)}\tilde{\boldsymbol{\mu}}_j^{(t)\top}\bigr)
    \tilde{\boldsymbol{\theta}}_i
    - \xi_{ij}^2
    \right\}
    \Bigg] \\
    &\qquad
    + \frac{J}{2}\log\!\left|\boldsymbol{\Sigma}_{\tilde{\boldsymbol{\beta}}}^{-1}\right|
    - \frac{1}{2}\sum_{j=1}^J
    \mathrm{tr}\!\left(
    \boldsymbol{\Sigma}_{\tilde{\boldsymbol{\beta}}}^{-1}
    \bigl(\tilde{\boldsymbol{\Sigma}}_j^{(t)}
    + \tilde{\boldsymbol{\mu}}_j^{(t)}\tilde{\boldsymbol{\mu}}_j^{(t)\top}\bigr)
    \right).
\end{align*}

\medskip
\noindent\textbf{M-step.}\\
The Q-function obtained in the E-step serves as the objective function to be maximized with respect to the ideal points $\{\theta_i\}$, the variational parameters $\{\xi_{ij}\}$ and the covariance matrix $\boldsymbol{\Sigma}_{\tilde{\boldsymbol{\beta}}}$.  
Since a joint maximization over all parameters does not lead to closed-form expressions, we adopt an alternating maximization scheme in which each parameter block is updated while the others are held fixed. The closed-form updates are as follows:

\medskip
\noindent \emph{Update of the ideal points.}\\
With $\xi_{ij}$ fixed at $\xi_{ij}^{(t)}$ and $\boldsymbol{\Sigma}_{\tilde{\boldsymbol{\beta}}}$ fixed at
$\boldsymbol{\Sigma}_{\tilde{\boldsymbol{\beta}}}^{(t)}$, the Q-function reduces to a function of $\{\theta_i\}$.
Maximizing this function with respect to $\{\theta_i\}$ yields the closed-form update
\begin{equation*}
\begin{split}
    \theta_i^{(t+1)}
    &=
    \frac{ \sum_{j=1}^J \Bigg\{
    \left(1/2 - y_{ij}\right)
    (\tilde{\boldsymbol{\mu}}_j^{(t)})_{(2)}
    -
    2\lambda(\xi_{ij}^{(t)})
    \left(\tilde{\boldsymbol{\Sigma}}_j^{(t)} 
    + \tilde{\boldsymbol{\mu}}_j^{(t)}\tilde{\boldsymbol{\mu}}_j^{(t)\top}\right)_{(1,2)}
    \Bigg\}}{2\sum_{j=1}^J 
    \lambda(\xi_{ij}^{(t)})
    \left(\tilde{\boldsymbol{\Sigma}}_j^{(t)} 
    + \tilde{\boldsymbol{\mu}}_j^{(t)}\tilde{\boldsymbol{\mu}}_j^{(t)\top}\right)_{(2,2)}},
\end{split}
\end{equation*}
where $(\cdot)_{(a)}$ and $(\cdot)_{(a,b)}$ denote vector components and matrix entries, respectively.

\medskip
\noindent \emph{Update of the variational parameters.}\\
After setting $\theta_i = \theta_i^{(t+1)}$ and holding $\boldsymbol{\Sigma}_{\tilde{\boldsymbol{\beta}}} = \boldsymbol{\Sigma}_{\tilde{\boldsymbol{\beta}}}^{(t)}$ fixed, the Q-function reduces to a function of $\{\xi_{ij}\}$. Maximizing this function with respect to $\{\xi_{ij}\}$ leads to the closed-form update
\begin{equation*}
    \xi_{ij}^{(t+1)}
    =
    \sqrt{
        \tilde{\boldsymbol{\theta}}_i^{(t+1)\top}
        \left(
            \tilde{\boldsymbol{\Sigma}}_j^{(t)}
            +
            \tilde{\boldsymbol{\mu}}_j^{(t)}
            \tilde{\boldsymbol{\mu}}_j^{(t)\top}
        \right)
        \tilde{\boldsymbol{\theta}}_i^{(t+1)}.
    }
\end{equation*}

\medskip
\noindent \emph{Update of the covariance matrix.}\\
Finally, fixing $\theta_i = \theta_i^{(t+1)}$ and $\xi_{ij} = \xi_{ij}^{(t+1)}$, the Q-function reduces to a function of $\boldsymbol{\Sigma}_{\tilde{\boldsymbol{\beta}}}$. Maximizing this function with respect to $\boldsymbol{\Sigma}_{\tilde{\boldsymbol{\beta}}}$ yields the closed-form update
\begin{equation*}
    \boldsymbol{\Sigma}_{\tilde{\boldsymbol{\beta}}}^{(t+1)} = \frac{1}{J} \sum_{j = 1}^J 
    \left(
            \tilde{\boldsymbol{\Sigma}}_j^{(t)}
            +
            \tilde{\boldsymbol{\mu}}_j^{(t)}
            \tilde{\boldsymbol{\mu}}_j^{(t)\top}
        \right).
\end{equation*}

\section{Implementation Details of the Variational Louis' Method}\label{appendix_B}
Our goal is to compute $\tilde{\mathcal{I}}_{\mathrm{obs}}(\hat{\boldsymbol{\vartheta}})$. By the first-order optimality condition of $\mathrm{ELBO}(q(\boldsymbol{z});\boldsymbol{\vartheta})$ with respect to $\boldsymbol{\vartheta}$,
\begin{equation*}
    \left.
    \mathbb{E}_{q(\boldsymbol{z})}\!\left[
    \nabla_{\boldsymbol{\vartheta}}
    \ell_e^{\mathrm{PG}}(\boldsymbol{\vartheta},\boldsymbol{z};\boldsymbol{y})
    \right]
    \right|_{\boldsymbol{\vartheta}=\hat{\boldsymbol{\vartheta}}}
    =0.
\end{equation*}
Thus, using $\mathrm{Var}(X)=\mathbb{E}[XX^\top]-\mathbb{E}[X]\mathbb{E}[X]^\top$, we obtain
\begin{align*}
    \tilde{\mathcal{I}}_{\mathrm{obs}}(\hat{\boldsymbol{\vartheta}})
    =
    -\left.
    \mathbb{E}_{q(\boldsymbol{z})}\!\left[
    \nabla_{\boldsymbol{\vartheta}}^2
    \ell_e^{\mathrm{PG}}(\boldsymbol{\vartheta},\boldsymbol{z};\boldsymbol{y})
    \right]
    \right|_{\boldsymbol{\vartheta}=\hat{\boldsymbol{\vartheta}}}
    -
    \left.
    \mathbb{E}_{q(\boldsymbol{z})}\!\left[
    \nabla_{\boldsymbol{\vartheta}}
    \ell_e^{\mathrm{PG}}(\boldsymbol{\vartheta},\boldsymbol{z};\boldsymbol{y})
    \nabla_{\boldsymbol{\vartheta}}^\top
    \ell_e^{\mathrm{PG}}(\boldsymbol{\vartheta},\boldsymbol{z};\boldsymbol{y})
    \right]
    \right|_{\boldsymbol{\vartheta}=\hat{\boldsymbol{\vartheta}}}.
\end{align*}

\subsection{Expected Hessian Term}
We first consider the expected Hessian term $\mathbb{E}_{q(\boldsymbol{z})}\!\left[\nabla_{\boldsymbol{\vartheta}}^2\ell_e^{\mathrm{PG}}(\boldsymbol{\vartheta},\boldsymbol{z};\boldsymbol{y})\right].$ Under the partition $\boldsymbol{\vartheta}=\{\boldsymbol{\Theta},\boldsymbol{\nu}\}$, it admits the block structure
\begin{equation*}
    \mathbb{E}_{q(\boldsymbol{z})}
    \left[
    \begin{pmatrix}
        \nabla_{\boldsymbol{\Theta}}^2\ell_e^{\mathrm{PG}}(\boldsymbol{\vartheta},\boldsymbol{z};\boldsymbol{y}) &
        \nabla_{\boldsymbol{\Theta}}\nabla_{\boldsymbol{\nu}}^\top\ell_e^{\mathrm{PG}}(\boldsymbol{\vartheta},\boldsymbol{z};\boldsymbol{y})\\
        \nabla_{\boldsymbol{\nu}}\nabla_{\boldsymbol{\Theta}}^\top\ell_e^{\mathrm{PG}}(\boldsymbol{\vartheta},\boldsymbol{z};\boldsymbol{y}) &
        \nabla_{\boldsymbol{\nu}}^2\ell_e^{\mathrm{PG}}(\boldsymbol{\vartheta},\boldsymbol{z};\boldsymbol{y})
    \end{pmatrix}
    \right].
\end{equation*}
For notational simplicity, we write $\ell_e^{\mathrm{PG}}(\boldsymbol{\vartheta}, \boldsymbol{z}; \boldsymbol{y})$ as $\ell_e^{\mathrm{PG}}$.

\medskip
\noindent\textbf{Block corresponding to $\boldsymbol{\Theta}$.}\\
The first-order derivative with respect to $\theta_i$ is
\begin{equation*}
    \frac{\partial \ell_e^{\mathrm{PG}}}{\partial\theta_i}
    =
    \sum_{j=1}^J
    \left\{
    \left(y_{ij}-\frac12\right)\beta_j
    -
    w_{ij}\bigl(\alpha_j\beta_j+\theta_i\beta_j^2\bigr)
    \right\}.
\end{equation*}
The second-order derivatives are
\begin{equation*}
    \frac{\partial^2\ell_e^{\mathrm{PG}}}{\partial\theta_i^2}
    =
    -\sum_{j=1}^J w_{ij}\beta_j^2,
    \qquad
    \frac{\partial^2\ell_e^{\mathrm{PG}}}{\partial\theta_i\partial\theta_{i'}}
    =0
    \quad(i\neq i').
\end{equation*}
Hence, $\mathbb{E}_{q(\boldsymbol{z})}\left[\nabla_{\boldsymbol{\Theta}}^2\ell_e^{\mathrm{PG}}\right]$ is diagonal, and its $k$th diagonal element is
\begin{equation*}
\begin{split}
    -\mathbb{E}_{q(\boldsymbol{z})}\!\left[\sum_{j=1}^J w_{kj}\beta_j^2\right]
    &=
    -\sum_{j=1}^J
    \mathbb{E}_{q_{kj}(w_{kj})}[w_{kj}]\,
    \mathbb{E}_{q_j(\tilde{\boldsymbol{\beta}}_j)}[\beta_j^2]\\
    &=
    -\sum_{j=1}^J
    \bar w_{kj}
    (\tilde{\boldsymbol{\Sigma}}_j+\tilde{\boldsymbol{\mu}}_j\tilde{\boldsymbol{\mu}}_j^\top)_{(2,2)},
\end{split}
\end{equation*}
where the first equality follows from the mean-field assumption.

\medskip
\noindent\textbf{Block corresponding to $\boldsymbol{\nu}$.}\\

Let $D := \sigma_{11}\sigma_{22} - \sigma_{12}^2$, and define
\[
    p_{11} := \frac{\sigma_{22}}{D}, \qquad
    p_{12} := -\frac{\sigma_{12}}{D}, \qquad
    p_{22} := \frac{\sigma_{11}}{D},
\]
i.e., $(p_{11},p_{12},p_{22})$ are the entries of the precision matrix
$\Sigma_{\tilde{\boldsymbol\beta}}^{-1}$, and let
\[
    S_{\alpha\alpha} := \sum_{j=1}^J \alpha_j^2, \qquad
    S_{\alpha\beta} := \sum_{j=1}^J \alpha_j\beta_j, \qquad
    S_{\beta\beta} := \sum_{j=1}^J \beta_j^2.
\]
The first-order derivatives with respect to each element of $\boldsymbol{\nu}$ are given
by
\begin{equation*}
\begin{split}
    \frac{\partial \ell_e^{\text{PG}}}{\partial \sigma_{11}}
    &= -\frac{J}{2}p_{11}
    + \frac{1}{2}\left(p_{11}^2 S_{\alpha\alpha} + 2p_{11}p_{12}S_{\alpha\beta} + p_{12}^2 S_{\beta\beta}\right), \\
    \frac{\partial \ell_e^{\text{PG}}}{\partial \sigma_{12}}
    &= -J p_{12}
    + \left(p_{11}p_{12} S_{\alpha\alpha} + (p_{11}p_{22}+p_{12}^2)S_{\alpha\beta} + p_{12}p_{22}S_{\beta\beta}\right), \\
    \frac{\partial \ell_e^{\text{PG}}}{\partial \sigma_{22}}
    &= -\frac{J}{2}p_{22}
    + \frac{1}{2}\left(p_{12}^2 S_{\alpha\alpha} + 2p_{12}p_{22}S_{\alpha\beta} + p_{22}^2 S_{\beta\beta}\right).
\end{split}
\end{equation*}
The second-order derivatives are
\begin{equation*}
\begin{split}
    \frac{\partial^2 \ell_e^{\text{PG}}}{\partial \sigma_{11}^2}
    &= \frac{J}{2}p_{11}^2
    - \left(p_{11}^3 S_{\alpha\alpha} + 2p_{11}^2p_{12}S_{\alpha\beta} + p_{11}p_{12}^2 S_{\beta\beta}\right), \\
    \frac{\partial^2 \ell_e^{\text{PG}}}{\partial \sigma_{11}\partial \sigma_{12}}
    &= Jp_{11}p_{12}
    - 2p_{11}^2p_{12}S_{\alpha\alpha}
    - p_{11}(p_{11}p_{22}+3p_{12}^2)S_{\alpha\beta}
    - p_{12}(p_{11}p_{22}+p_{12}^2)S_{\beta\beta}, \\
    \frac{\partial^2 \ell_e^{\text{PG}}}{\partial \sigma_{11}\partial \sigma_{22}}
    &= \frac{J}{2}p_{12}^2
    - p_{11}p_{12}^2 S_{\alpha\alpha}
    - p_{12}(p_{11}p_{22}+p_{12}^2)S_{\alpha\beta}
    - p_{12}^2 p_{22} S_{\beta\beta}, \\
    \frac{\partial^2 \ell_e^{\text{PG}}}{\partial \sigma_{12}^2}
    &= J(p_{11}p_{22}+p_{12}^2)
    - p_{11}(p_{11}p_{22}+3p_{12}^2)S_{\alpha\alpha}
    - 2p_{12}(3p_{11}p_{22}+p_{12}^2)S_{\alpha\beta}
    - p_{22}(p_{11}p_{22}+3p_{12}^2)S_{\beta\beta}, \\
    \frac{\partial^2 \ell_e^{\text{PG}}}{\partial \sigma_{12}\partial \sigma_{22}}
    &= Jp_{12}p_{22}
    - p_{12}(p_{11}p_{22}+p_{12}^2)S_{\alpha\alpha}
    - p_{22}(p_{11}p_{22}+3p_{12}^2)S_{\alpha\beta}
    - 2p_{12}p_{22}^2 S_{\beta\beta}, \\
    \frac{\partial^2 \ell_e^{\text{PG}}}{\partial \sigma_{22}^2}
    &= \frac{J}{2}p_{22}^2
    - p_{12}^2 p_{22} S_{\alpha\alpha} - 2p_{12}p_{22}^2 S_{\alpha\beta} - p_{22}^3 S_{\beta\beta}.
\end{split}
\end{equation*}
Since the second-order derivatives do not depend on the latent variables $w_{ij}$,
\begin{equation}\label{eq:block_nu_1}
\begin{split}
    \mathbb{E}_{q(\boldsymbol{z})}\!\left[
    \nabla_{\boldsymbol{\nu}}^2\ell_e^{\mathrm{PG}}
    \right]
    =
    \mathbb{E}_{q(\boldsymbol{B})}\!\left[\nabla_{\boldsymbol{\nu}}^2\ell_e^{\mathrm{PG}}\right]
    =
    \mathbb{E}_{q(\boldsymbol{B})}\!\left[
    \begin{pmatrix}
        \frac{\partial^2 \ell_e^{\mathrm{PG}}}{\partial \sigma_{11}^2} &
        \frac{\partial^2 \ell_e^{\mathrm{PG}}}{\partial \sigma_{11}\partial \sigma_{12}} &
        \frac{\partial^2 \ell_e^{\mathrm{PG}}}{\partial \sigma_{11}\partial \sigma_{22}}\\[6pt]
        \frac{\partial^2 \ell_e^{\mathrm{PG}}}{\partial \sigma_{12}\partial \sigma_{11}} &
        \frac{\partial^2 \ell_e^{\mathrm{PG}}}{\partial \sigma_{12}^2} &
        \frac{\partial^2 \ell_e^{\mathrm{PG}}}{\partial \sigma_{12}\partial \sigma_{22}}\\[6pt]
        \frac{\partial^2 \ell_e^{\mathrm{PG}}}{\partial \sigma_{22}\partial \sigma_{11}} &
        \frac{\partial^2 \ell_e^{\mathrm{PG}}}{\partial \sigma_{22}\partial \sigma_{12}} &
        \frac{\partial^2 \ell_e^{\mathrm{PG}}}{\partial \sigma_{22}^2}
    \end{pmatrix}\right].
\end{split}
\end{equation}

\medskip
\noindent\textbf{Cross block.}\\
Since $\partial \ell_e^{\mathrm{PG}}/\partial\theta_i$ does not depend on $\boldsymbol{\nu}$ for all $i$, we have $\nabla_{\boldsymbol{\Theta}}\nabla_{\boldsymbol{\nu}}^\top \ell_e^{\mathrm{PG}} = \mathbf{0}$, and thereby 
\begin{equation*}
    \mathbb{E}_{q(\boldsymbol{z})}\left[\nabla_{\boldsymbol{\Theta}}\nabla_{\boldsymbol{\nu}}^\top
    \ell_e^{\mathrm{PG}}\right]=
    \mathbf{0}.
\end{equation*}

\subsection{Expected Outer Product of Gradients Term}
We next consider the expected outer product of gradients term $\mathbb{E}_{q(\boldsymbol{z})}\!\left[\nabla_{\boldsymbol{\vartheta}}\ell_e^{\mathrm{PG}}\nabla_{\boldsymbol{\vartheta}}^\top\ell_e^{\mathrm{PG}}\right].$ Under the partition $\boldsymbol{\vartheta}=\{\boldsymbol{\Theta},\boldsymbol{\nu}\}$, this term has block structure
\begin{equation*}
    \mathbb{E}_{q(\boldsymbol{z})}
    \left[
    \begin{pmatrix}
        \nabla_{\boldsymbol{\Theta}}\ell_e^{\mathrm{PG}}
        \nabla_{\boldsymbol{\Theta}}^\top\ell_e^{\mathrm{PG}}
        &
        \nabla_{\boldsymbol{\Theta}}\ell_e^{\mathrm{PG}}
        \nabla_{\boldsymbol{\nu}}^\top\ell_e^{\mathrm{PG}}\\
        \nabla_{\boldsymbol{\nu}}\ell_e^{\mathrm{PG}}
        \nabla_{\boldsymbol{\Theta}}^\top\ell_e^{\mathrm{PG}}
        &
        \nabla_{\boldsymbol{\nu}}\ell_e^{\mathrm{PG}}
        \nabla_{\boldsymbol{\nu}}^\top\ell_e^{\mathrm{PG}}
    \end{pmatrix}
    \right].
\end{equation*}
For notational simplicity, let $\kappa_{ij} = y_{ij} - 1/2$, and let 
$q(\boldsymbol{B}) = \prod_{j} q_j(\tilde{\boldsymbol{\beta}}_j)$ and 
$q_k(\boldsymbol{w_{k,\cdot}}) = \prod_j q_{kj}(w_{kj})$.

\medskip
\noindent\textbf{Block corresponding to $\boldsymbol{\Theta}$.}\\
Define $v_{ij}=\mathrm{Var}_{q_{ij}(w_{ij})}(w_{ij})$. The $(k,l)$th element of $\mathbb{E}_{q(\boldsymbol{z})}\!\left[\nabla_{\boldsymbol{\Theta}}\ell_e^{\mathrm{PG}}\nabla_{\boldsymbol{\Theta}}^\top\ell_e^{\mathrm{PG}}\right]$ is given by
\begin{equation}
\begin{split}\label{eq:block_theta_off}
    &\left(
    \mathbb{E}_{q(\boldsymbol{z})}\!\left[
    \nabla_{\boldsymbol{\Theta}} \ell_e^{\mathrm{PG}}
    \nabla_{\boldsymbol{\Theta}}^{\top} \ell_e^{\mathrm{PG}}
    \right]
    \right)_{(k,l)} \\
    &\quad=
    \mathbb{E}_{q(\boldsymbol{B})}\!\left[
    \left\{
    \sum_{j=1}^{J}
    \Bigl(
    \kappa_{kj}\beta_j - \bar{w}_{kj}(\alpha_j\beta_j + \theta_k\beta_j^2)
    \Bigr)
    \right\}
    \left\{
    \sum_{j=1}^{J}
    \Bigl(
    \kappa_{lj}\beta_j - \bar{w}_{lj}(\alpha_j\beta_j + \theta_l\beta_j^2)
    \Bigr)
    \right\}
    \right],
\end{split}
\end{equation}
for $k \neq l$, and for $k = l$
\begin{equation}\label{eq:block_theta_diag}
\begin{split}
    &\left(
    \mathbb{E}_{q(\boldsymbol{z})}\!\left[
    \nabla_{\boldsymbol{\Theta}} \ell_e^{\mathrm{PG}}
    \nabla_{\boldsymbol{\Theta}}^{\top} \ell_e^{\mathrm{PG}}
    \right]
    \right)_{(k,k)}\\
    &\quad=
    \mathbb{E}_{q(\boldsymbol{B})}\!\left[
    \left\{
    \sum_{j=1}^{J}
    \Bigl(
    \kappa_{kj}\beta_j - \bar{w}_{kj}(\alpha_j\beta_j + \theta_k\beta_j^2)
    \Bigr)
    \right\}^2
    +
    \sum_{j=1}^{J}
    v_{kj}(\alpha_j\beta_j + \theta_k\beta_j^2)^2
    \right].
\end{split}
\end{equation}
The equality follows from the law of iterated expectations, combined with the mean-field factorization.

\medskip
\noindent\textbf{Block corresponding to $\boldsymbol{\nu}$.}\\
Writing
\begin{equation*}
\begin{split}
    \nabla_{\boldsymbol{\nu}}\ell_e^{\mathrm{PG}}
    =
    \left(
    \frac{\partial \ell_e^{\mathrm{PG}}}{\partial \nu_{1}},\;
    \frac{\partial \ell_e^{\mathrm{PG}}}{\partial \nu_{2}},\;
    \frac{\partial \ell_e^{\mathrm{PG}}}{\partial \nu_{3}}
    \right)^\top
    =
    \left(
    \frac{\partial \ell_e^{\mathrm{PG}}}{\partial \sigma_{11}},\;
    \frac{\partial \ell_e^{\mathrm{PG}}}{\partial \sigma_{12}},\;
    \frac{\partial \ell_e^{\mathrm{PG}}}{\partial \sigma_{22}}
    \right)^\top,
\end{split}
\end{equation*}
we have
\begin{equation}\label{eq:block_nu_2}
\begin{split}
    \mathbb{E}_{q(\boldsymbol{z})}\!\left[
    \nabla_{\boldsymbol{\nu}}\ell_e^{\mathrm{PG}}
    \nabla_{\boldsymbol{\nu}}^\top\ell_e^{\mathrm{PG}}
    \right]
    =
    \mathbb{E}_{q(\boldsymbol{B})}\!\left[
    \begin{pmatrix}
        \left(\frac{\partial \ell_e^{\mathrm{PG}}}{\partial \nu_{1}}\right)^2 &
        \frac{\partial \ell_e^{\mathrm{PG}}}{\partial \nu_{1}}
        \frac{\partial \ell_e^{\mathrm{PG}}}{\partial \nu_{2}} &
        \frac{\partial \ell_e^{\mathrm{PG}}}{\partial \nu_{1}}
        \frac{\partial \ell_e^{\mathrm{PG}}}{\partial \nu_{3}} \\
        \frac{\partial \ell_e^{\mathrm{PG}}}{\partial \nu_{2}}
        \frac{\partial \ell_e^{\mathrm{PG}}}{\partial \nu_{1}} &
        \left(\frac{\partial \ell_e^{\mathrm{PG}}}{\partial \nu_{2}}\right)^2 &
        \frac{\partial \ell_e^{\mathrm{PG}}}{\partial \nu_{2}}
        \frac{\partial \ell_e^{\mathrm{PG}}}{\partial \nu_{3}} \\
        \frac{\partial \ell_e^{\mathrm{PG}}}{\partial \nu_{3}}
        \frac{\partial \ell_e^{\mathrm{PG}}}{\partial \nu_{1}} &
        \frac{\partial \ell_e^{\mathrm{PG}}}{\partial \nu_{3}}
        \frac{\partial \ell_e^{\mathrm{PG}}}{\partial \nu_{2}} &
        \left(\frac{\partial \ell_e^{\mathrm{PG}}}{\partial \nu_{3}}\right)^2
    \end{pmatrix}
    \right]. 
\end{split}
\end{equation}
The equality follows from the fact that $\partial \ell_e^{\mathrm{PG}}/\partial \nu_i$ does not depend on $\boldsymbol{w}$ for all $i$, combined with the mean-field factorization.

\medskip
\noindent\textbf{Cross block.}\\
The $(k,l)$th entry of $\mathbb{E}_{q(\boldsymbol{z})}\!\left[ \nabla_{\boldsymbol{\Theta}}\ell_e^{\mathrm{PG}} \nabla_{\boldsymbol{\nu}}^\top\ell_e^{\mathrm{PG}} \right]$ is given by
\begin{align}\label{eq:block_cross}\nonumber
    \left(
    \mathbb{E}_{q(\boldsymbol{z})}\!\left[
    \nabla_{\boldsymbol{\Theta}}\ell_e^{\mathrm{PG}}
    \nabla_{\boldsymbol{\nu}}^\top\ell_e^{\mathrm{PG}}
    \right]
    \right)_{(k,l)}
    &= \mathbb{E}_{q(\boldsymbol{B})}\!
    \left[ \mathbb{E}_{q_k(\boldsymbol{w_{k,\cdot}})}
    \left\{\frac{\partial \ell_e^{\mathrm{PG}}}{\partial \theta_k}
    \frac{\partial \ell_e^{\mathrm{PG}}}{\partial \nu_l}\right\}
    \right]\\ 
    &= \mathbb{E}_{q(\boldsymbol{B})}\!
    \left\{
    \sum_{j=1}^J
    \Bigl(
    \kappa_{kj}\beta_j- \bar{w}_{kj}(\alpha_j\beta_j+\theta_k\beta_j^2)
    \Bigr)
    \frac{\partial \ell_e^{\mathrm{PG}}}{\partial \nu_l}\right\}
\end{align}
The first equality follows from the mean-field assumption, and the second equality follows from the fact that $\partial \ell_e^{\mathrm{PG}}/\partial \nu_l$ does not depend on the latent variable $\boldsymbol{w}$.

Note that all four terms in \eqref{eq:block_nu_1}--\eqref{eq:block_cross} depend on expectations taken with respect to $q(\boldsymbol{B}) = \prod_j q_j(\tilde{\boldsymbol{\beta}}_j)$. Although these expectations can be evaluated analytically, the resulting expressions are cumbersome \citep{kan2008moments}; we therefore propose using a Monte Carlo approximation in practice. Specifically, we draw a sufficiently large number of samples from $q(\boldsymbol{z})$ and approximate the expectations via the law of large numbers.

Combining all the above results, we can obtain $\tilde{\mathcal{I}}_{\mathrm{obs}}(\hat{\boldsymbol{\vartheta}})$. The standard error of $\hat{\theta}_i$ is then given by
\begin{equation*}
    \mathrm{se}(\hat{\theta}_i)=
    \left\{
    \left(
    \tilde{\mathcal{I}}_{\boldsymbol{\Theta}}(\boldsymbol{\hat{\vartheta}})
    -
    \tilde{\mathcal{I}}_{\boldsymbol{\Theta}, \boldsymbol{\nu}}(\boldsymbol{\hat{\vartheta}})
    \tilde{\mathcal{I}}_{\boldsymbol{\nu}}^{-1}(\boldsymbol{\hat{\vartheta}})
    \tilde{\mathcal{I}}_{\boldsymbol{\Theta}, \boldsymbol{\nu}}^\top(\boldsymbol{\hat{\vartheta}})
    \right)^{-1}
    _{(i,i)}
    \right\}^{1/2}.
\end{equation*}


\section{An Additional Real Data Analysis: the 118th U.S. Congress}\label{appendix_C}
\begin{figure}[!htb]
    \centering
    \includegraphics[width=\linewidth]{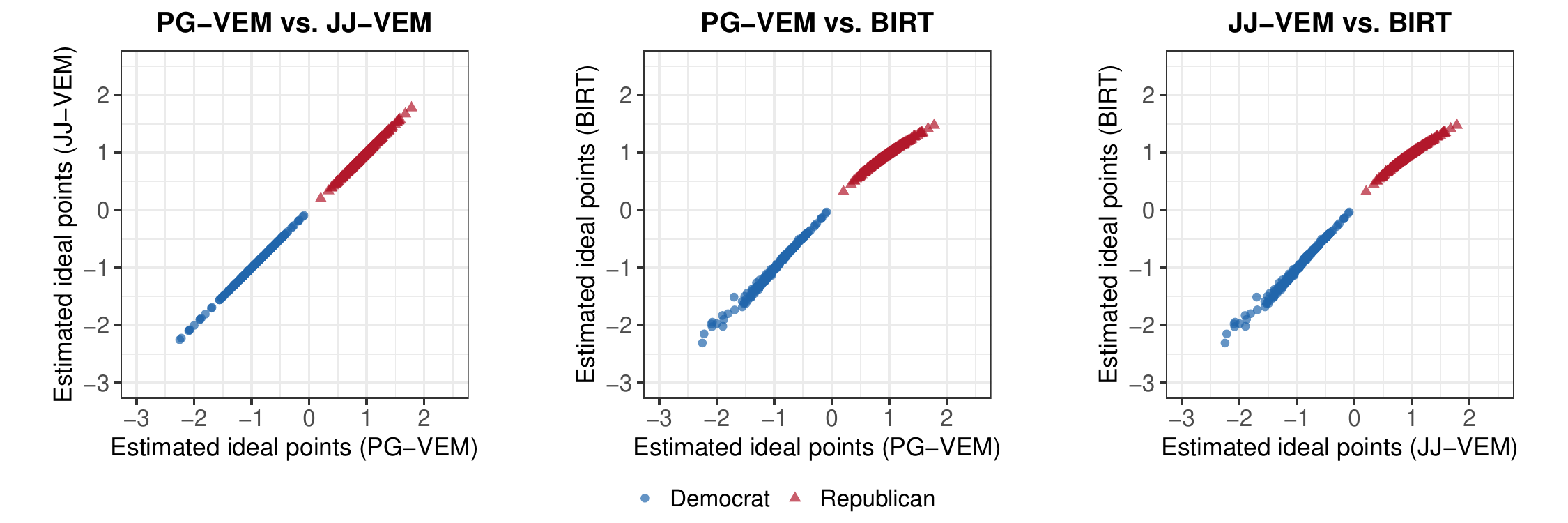}
    \caption{The 118th U.S. Congress: comparison of the estimated ideal points obtained by \textsc{PG-VEM}, \textsc{JJ-VEM}, and \textsc{BIRT}. Blue circles denote Democrats, and red triangles denote Republicans.}
    \label{fig:13}
\end{figure}

\begin{figure}[!htb]
    \centering
    \includegraphics[width=0.6\linewidth]{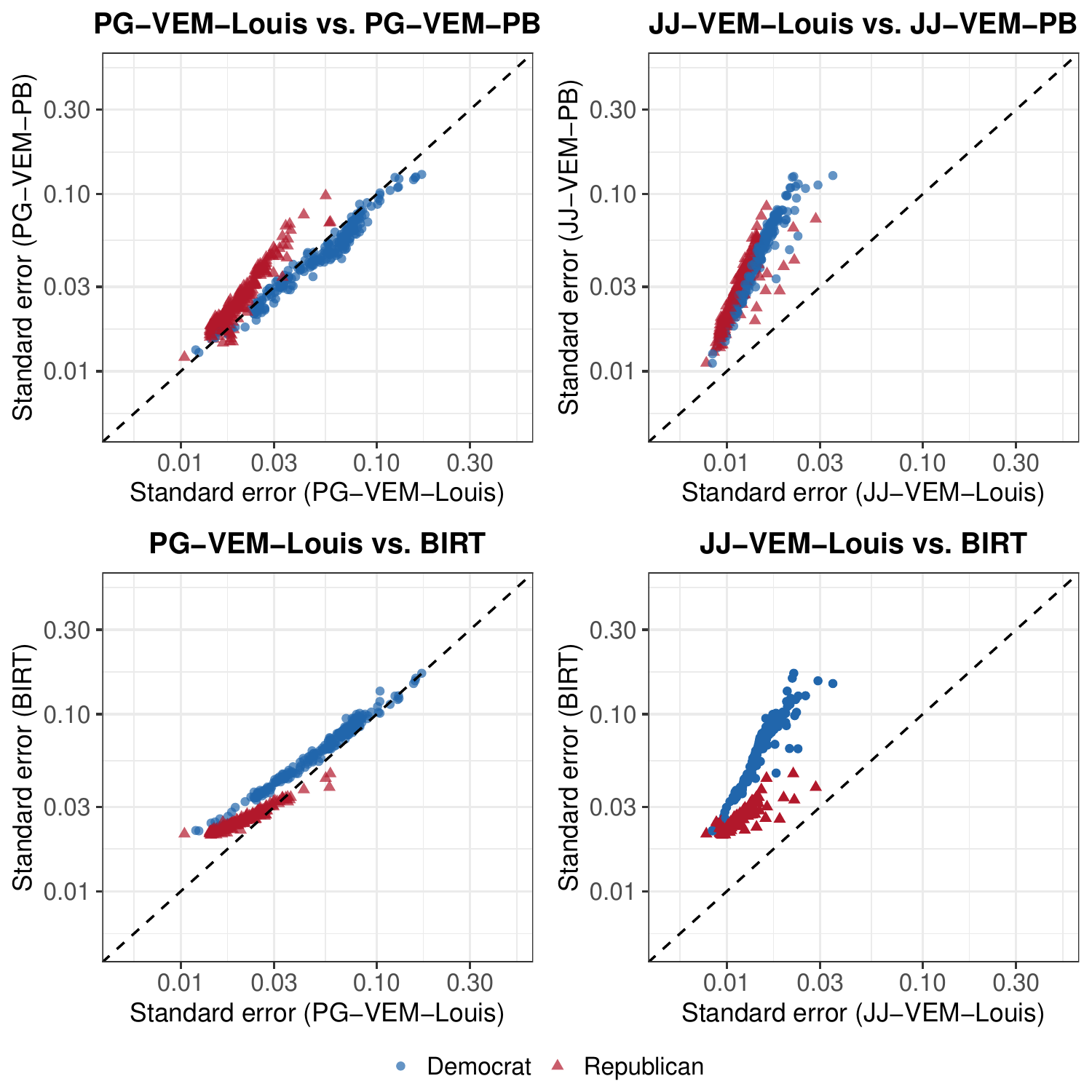}
    \caption{The 118th U.S. Congress: comparison of standard errors for the estimated ideal points under five methods---\textsc{PG-VEM-PB}, \textsc{PG-VEM-Louis}, \textsc{JJ-VEM-PB}, \textsc{JJ-VEM-Louis}, and \textsc{BIRT}. In each panel, both axes are on a $\log_{10}$ scale, and the black dashed line represents the 45-degree line. Blue circles indicate Democrats, and red triangles indicate Republicans.}
    \label{fig:14}
\end{figure}

\begin{figure}[!htb]
    \centering
    \includegraphics[width=\linewidth]{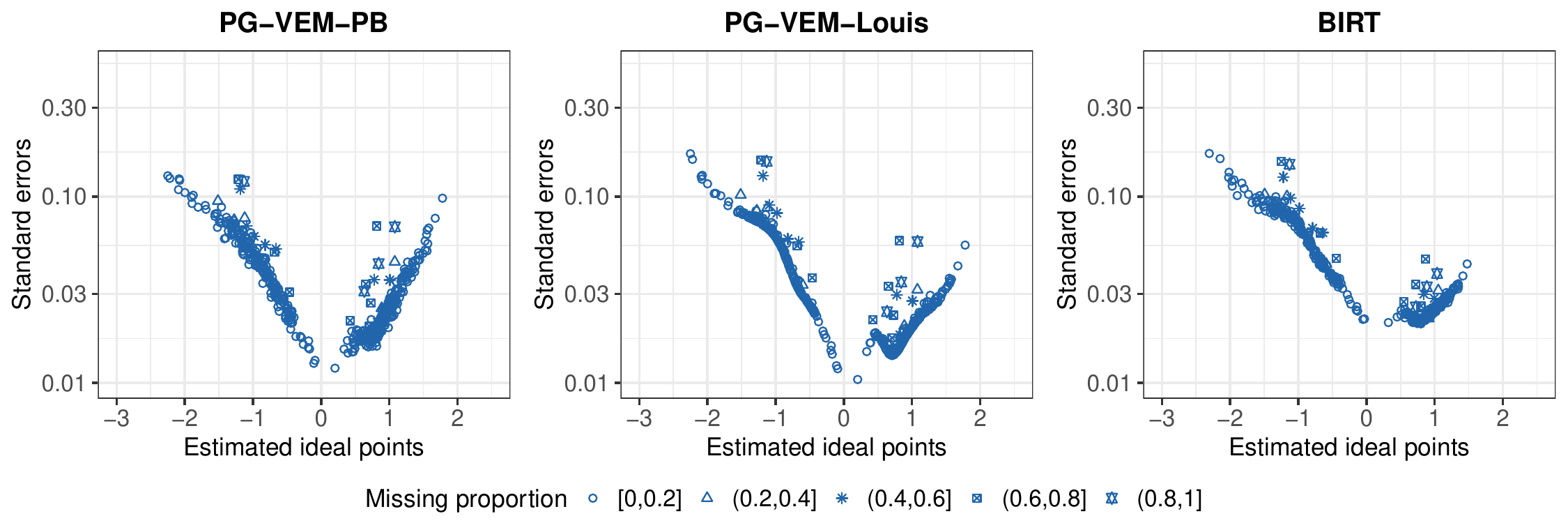}
    \caption{The 118th U.S. Congress: scatter plots of estimated ideal points (horizontal axis) versus estimated standard errors (vertical axis, on the $\log_{10}$ scale), with point shapes indicating the missing rate. From left to right, the three panels correspond to the results from \textsc{PG-VEM-PB}, \textsc{PG-VEM-Louis}, and \textsc{BIRT}.}
    \label{fig:15}
\end{figure}

In this section, we apply \textsc{PG-VEM}, \textsc{JJ-VEM}, and \textsc{BIRT} to roll-call data from the 118th U.S.\ Congress. The results are summarized in Figures~\ref{fig:13}--\ref{fig:15}. We observe patterns consistent with those in the numerical studies and in the analysis of roll-call data from the 113th U.S.\ Congress presented in the main manuscript.

\section{Additional Simulation Results: Dependence Among Bill Parameters}
\label{appendix_D}

To examine the sensitivity of \textsc{PG-VEM} and \textsc{PG-VEM-Louis} to a violation of the mean-field assumption, we consider a variant of Simulation Scenario~I in which the bill parameters are no longer independent across bills. All aspects of the design follow Scenario~I except for the generation of the bill parameters $\tilde{\boldsymbol{\beta}}_j = (\alpha_j,\beta_j)$: rather than drawing each $\tilde{\boldsymbol{\beta}}_j$ independently from $N_2(\mathbf{0},\Sigma_{\tilde{\boldsymbol{\beta}}})$, we partition the $J$ bills into consecutive, non-overlapping groups of a specified size and draw a single value of $\tilde{\boldsymbol{\beta}}_j$ from $N_2(\mathbf{0},\Sigma_{\tilde{\boldsymbol{\beta}}})$ for each group, which is then shared by every bill within that group. For instance, with a group size of $4$, bills $1$--$4$ share one draw of $\tilde{\boldsymbol{\beta}}_j$, bills $5$--$8$ share an independent second draw, and so on. This construction leaves the marginal distribution of each $\tilde{\boldsymbol{\beta}}_j$ unchanged (still $N_2(\mathbf{0},\Sigma_{\tilde{\boldsymbol{\beta}}})$) but induces perfect within-group correlation. A group size of $1$ recovers the original, independent design of Scenario~I; we consider a group size of $4$ here to evaluate the impact of this dependence on the resulting ideal point estimates and standard errors.

\begin{figure}[!htb]
    \centering
    \includegraphics[width=0.7\linewidth]{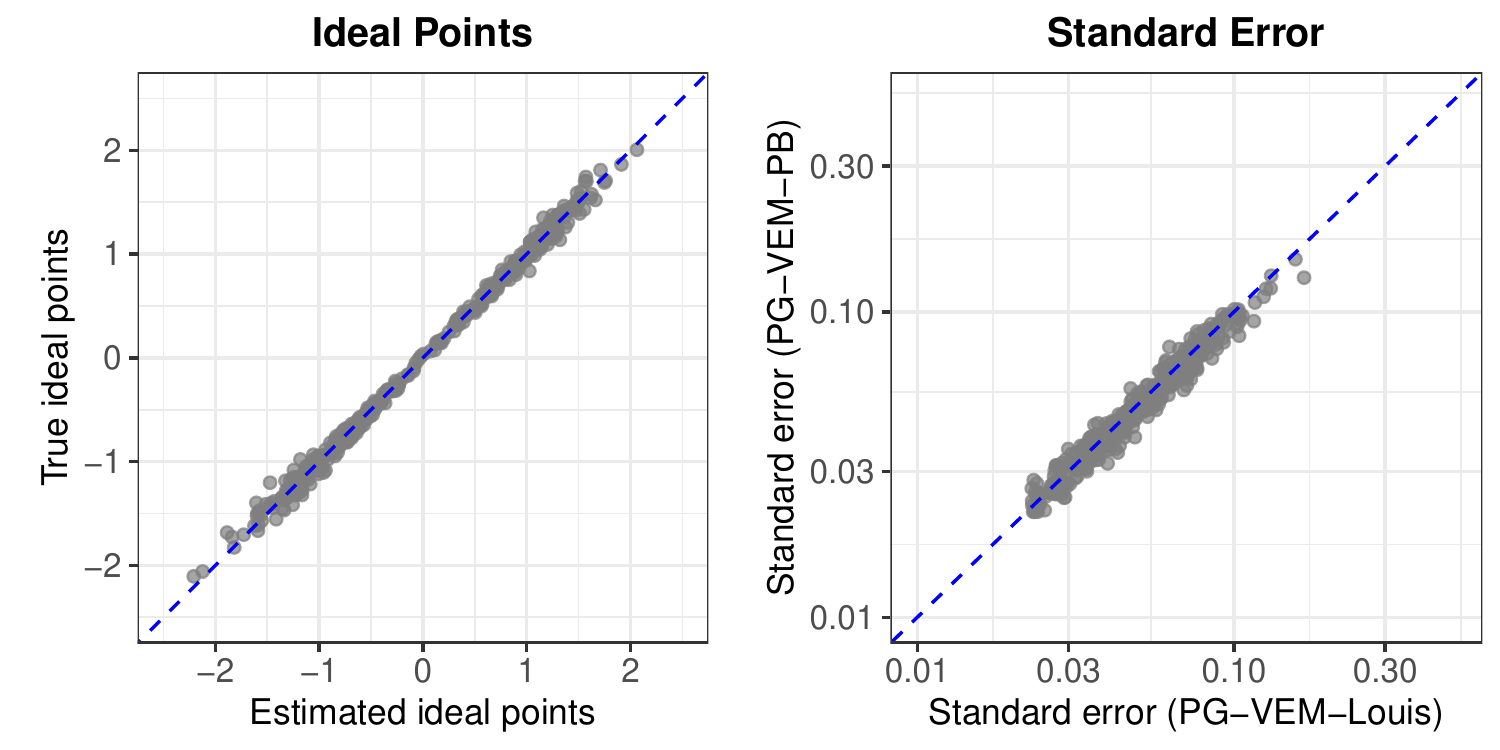}
    \caption{Effect of induced bill-parameter dependence (group size $4$) on \textsc{PG-VEM} and \textsc{PG-VEM-Louis}. Left: estimated ideal points via \textsc{PG-VEM} versus their true values. Right: standard errors from \textsc{PG-VEM-Louis} versus \textsc{PG-VEM-PB}, both on a $\log_{10}$ scale. In both panels, the blue dashed line indicates the 45-degree line.}
    \label{fig:16}
\end{figure}

Figure~\ref{fig:16} summarizes the results. The left panel compares the estimated ideal points from \textsc{PG-VEM} against their true values; despite the induced within-group dependence, the estimates remain close to the 45-degree line, comparable to the independent case in Scenario~I. The right panel compares the standard errors from \textsc{PG-VEM-Louis} against those from \textsc{PG-VEM-PB} (based on 100 bootstrap replicates); the two sets of standard errors continue to closely track each other, with no visible degradation relative to the independent design. 

Taken together, these results indicate that, at least for this particular form and degree of dependence, the violation of the mean-field assumption does not noticeably affect the accuracy of \textsc{PG-VEM} or \textsc{PG-VEM-Louis}. However, we emphasize that this finding is based on a single simulation setting, and generalizing beyond this specific form and degree of dependence should be done with caution.

\section{A More Detailed Explanation of the Spatial Voting Model}\label{appendix_E}
In this section, we provide a more detailed account of the spatial voting model introduced in Section~\ref{sub_sec_2.1}.

To explain legislators' voting behavior in an interpretable manner, we adopt a utility-based spatial voting model that links observed votes to latent policy preferences \citep{enelow1984spatial}. We first define legislators' policy preferences in terms of utilities, and then explain how these utilities are connected to the observed vote.

Specifically, suppose that there exists a latent policy space, represented as a one-dimensional Euclidean space, $\mathbb{R}$. Let $\theta_i \in \mathbb{R}$ denote the ideological position of legislator $i$ in this policy space, commonly referred to as the ideal point, and let $\zeta_j, \psi_j \in \mathbb{R}$ denote the policy positions associated with the yea and nay outcomes of bill $j$, respectively, both of which also reside in this same policy space. Given $\theta_i$, $\zeta_j$, and $\psi_j$, legislator $i$'s decision to cast a yea or nay vote on bill $j$ is governed by utilities associated with voting yea and nay, each consisting of two parts: a deterministic part, which reflects how close $\theta_i$ is to $\zeta_j$ and $\psi_j$, respectively, and a stochastic part, which absorbs any remaining, unobserved influences on the vote. 

We model the deterministic part using a quadratic specification, as in \citet{poole2001geometry} and \citet{clinton2004statistical}, and the stochastic part using Gumbel-distributed errors, as in \citet{poole1985spatial}. Under these choices, the utilities associated with voting yea and nay are given by
\begin{equation*}
    U_{ij}^{\mathrm{yea}} 
    = \underbrace{-(\theta_i-\zeta_j)^2}_{\text{Deterministic}} \;\;+ \underbrace{\eta_{ij}}_{\text{Stochastic}},
    \quad 
    U_{ij}^{\mathrm{nay}} 
    = \underbrace{-(\theta_i-\psi_j)^2}_{\text{Deterministic}} \;\;+ \underbrace{\nu_{ij}}_{\text{Stochastic}},
\end{equation*}
where $\eta_{ij}$ and $\nu_{ij}$ are independent stochastic error terms following the Gumbel distribution with common location parameter $0$ and common scale parameter $\sigma_j$. The utility of legislator $i$ for bill $j$ is then defined as the difference
\[
    U_{ij} = U_{ij}^{\mathrm{yea}} - U_{ij}^{\mathrm{nay}} =  \underbrace{- (\theta_i - \zeta_j)^2 + (\theta_i - \psi_j)^2}_{\text{Deterministic}} + \underbrace{(\eta_{ij} - \nu_{ij})}_{\text{Stochastic}},
\]
which reflects legislator $i$'s policy preference on bill $j$: for example, $U_{ij} > 0$ means that the utility of voting yea exceeds that of voting nay, so legislator $i$ prefers to cast a yea vote; conversely, $U_{ij} < 0$ means that the utility of voting nay exceeds that of voting yea, so legislator $i$ prefers to cast a nay vote.

It remains to connect these utilities $U_{ij}$ to the observed votes $Y_{ij}$. We assume that legislator $i$ votes yea on bill $j$ whenever $U_{ij} \ge 0$, and votes nay otherwise. It then follows that
\begin{equation*} 
\begin{split}
    \mathbb{P}\!\left(Y_{ij}=1\right) 
    =
    \mathbb{P}\!\left(U_{ij}\ge 0 \right)
    =
    \mathbb{P}\!\left( X_{ij} \le \frac{\psi_j^2-\zeta_j^2}{\sigma_j}+\frac{2(\zeta_j-\psi_j)}{\sigma_j}\,\theta_i \right) 
    =
    \sigma(\alpha_j+\beta_j\theta_i), 
\end{split}
\end{equation*} 
where $X_{ij}$ is a standard logistic random variable, $\sigma(\cdot)$ denotes its distribution function given by $\sigma(x) = \{1 + \exp(-x)\}^{-1}$, $\alpha_j = (\psi_j^2 - \zeta_j^2)/\sigma_j$, and $\beta_j = 2(\zeta_j - \psi_j)/\sigma_j$. The second equality follows from the fact that $\eta_{ij}-\nu_{ij} \sim \mathrm{Logistic}(0,\sigma_j)$, where $\mathrm{Logistic}(\mu,\sigma)$ denotes the logistic distribution with location parameter $\mu$ and scale parameter $\sigma$; rearranging terms and simplifying yields the stated expression.

\end{document}